\documentclass[superscriptaddress,
amsmath,amssymb,
aps,prb,floatfix,twocolumn,10pt
]{revtex4-2}

\usepackage[utf8]{inputenc}
\usepackage{amsmath}
\usepackage{amssymb}
\usepackage{graphicx}
\usepackage{hyperref}
\usepackage{braket}
\usepackage{babel}
\usepackage{mathtools}
\usepackage{xspace}
\usepackage{xcolor}
\usepackage{physics}
\usepackage{comment}

\begin{document}

\title{Theory of phonon spectroscopy with the quantum twisting microscope}

\author{Jiewen Xiao}
\affiliation{\mbox{Department of Condensed Matter Physics, Weizmann Institute of Science, Rehovot 76100, Israel}}

\author{Erez Berg}
\affiliation{\mbox{Department of Condensed Matter Physics, Weizmann Institute of Science, Rehovot 76100, Israel}}

\author{Leonid I.\ Glazman}
\affiliation{\mbox{Department of Physics and Yale Quantum Institute, Yale University, New Haven, Connecticut 06520, USA}}

\author{Francisco Guinea}
\affiliation{\mbox{Imdea Nanoscience, Faraday 9, 28015 Madrid, Spain}}
\affiliation{\mbox{Donostia International Physics Center, Paseo Manuel de Lardiz\'abal 4, 20018 San Sebastian, Spain}}

\author{Shahal Ilani}
\affiliation{\mbox{Department of Condensed Matter Physics, Weizmann Institute of Science, Rehovot 76100, Israel}}

\author{Felix von Oppen}
\affiliation{\mbox{Dahlem Center for Complex Quantum Systems and Fachbereich Physik, Freie Universit\"at Berlin, 14195 Berlin, Germany}}

\date{\today}

\begin{abstract}
We develop a theory of probing phonon modes of van-der-Waals materials using the quantum twisting microscope. While elastic tunneling dominates the tunneling current at small twist angles, the momentum mismatch between the $K$ points of tip and sample at large twist angles can only be bridged by inelastic scattering. This allows for probing phonon dispersions along certain lines in reciprocal space by measuring the tunneling current as a function of twist angle and bias voltage. We illustrate this modality of the quantum twisting microscope by developing a systematic theory for  graphene-graphene junctions. We show that beyond phonon dispersions, the tunneling current also encodes the strength of electron-phonon couplings. Extracting the coupling strengths for individual phonon modes requires careful consideration of various inelastic tunneling processes. These processes are associated with the intralayer and interlayer electron-phonon couplings and appear at different orders in a perturbative calculation of the tunneling current.  
We find that the dominant process depends on the particular phonon mode under consideration. Our results inform the quest to understand the origin of superconductivity in twisted bilayer graphene and provide a case study for quantum-twisting-microscope investigations of collective modes. 
\end{abstract}

\maketitle

\section{Introduction}

Beyond conventional transport experiments, much information on van-der-Waals materials derives from local probes such as scanning tunneling microscopy (STM) as well as scanning single-electron-transistor and scanning SQUID measurements. A particularly well-studied system is twisted bilayer graphene with its multitude of electronic phases when the twist angle is close to the magic angle \cite{Cao2018a,Cao2018b,Andrei2021}. Scanning tunneling microscopy has been instrumental in elucidating the effects of correlations on their flat-band dispersion \cite{Choi2019,Kerelsky2019,Xie2019,Jiang2019}, of superconductivity \cite{Oh2021}, and of the correlated insulating states \cite{Nuckolls2023}. Compressibility measurements using a scanning single-electron transistor revealed a cascade of flavor-polarized phases \cite{Zondiner2020,Wong2020} as well as excess entropy associated with the formation of local moments \cite{Rozen2021}. It was also instrumental in uncovering anomalous Chern-insulator phases \cite{Xie2021}. Scanning-SQUID measurements revealed the formation of Chern mosaics \cite{Grover2022}. 

\begin{figure}[b]
\includegraphics[width=.9\linewidth]{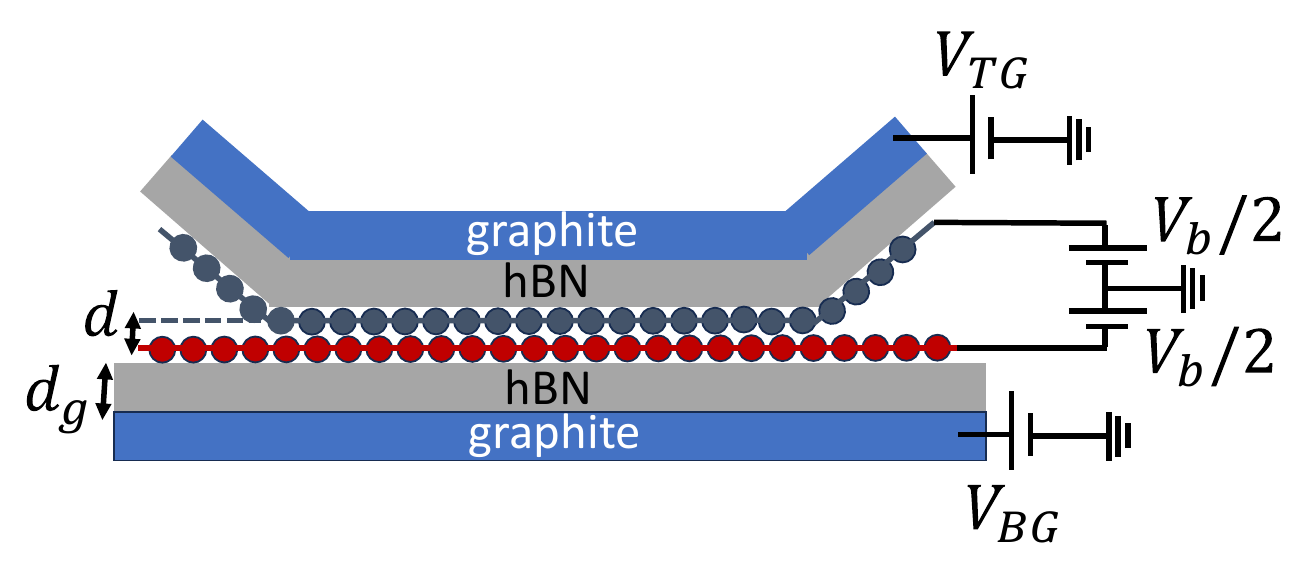}
\caption{Schematic setup of QTM with scanning tip (top) and fixed sample (bottom). The electron densities of the van-der-Waals layers (here: monolayer graphene) on tip and substrate are controlled by independent top (TG) and bottom (BG) gates (here: graphite). The gate electrodes are separated from the van-der-Waals layers by a gate dielectric (here: hBN).
For elastic-tunneling experiments at small twist angles, tip and sample are separated by an additional barrier layer (e.g., WSe$_2$; not shown). No additional barrier is needed when probing inelastic tunneling currents at larger twist angles.}
\label{fig:elst}
\end{figure}

The quantum twisting microscope (QTM) \cite{Inbar2023} is a powerful new instrument complementing previously existing local probes. Rather than measuring the local tunneling current at the atomic scale as in scanning tunneling microscopy, it relies on coherent tunneling across a twistable finite-area junction formed at the interface between van-der-Waals systems placed on a scanning tip with a flat pyramidal top and on a substrate, see Fig.\ \ref{fig:elst}. Due to the finite contact area, tunneling conserves crystal momentum modulo reciprocal lattice vectors of the tip and sample layers. Except at special twist angles, umklapp processes involving larger reciprocal lattice vectors will typically be suppressed. The twist imposes a relative rotation of the dispersions of tip and sample in momentum space. Moreover, the bias voltage introduces a relative shift of the dispersions in energy. Measuring the tunneling current as a function of bias voltage and twist angle will then provide direct signatures of momentum-resolved dispersions. This has been used to explore the electronic dispersion of graphene layers using graphene-graphene junctions as well as the flat-band dispersions of twisted bilayer graphene using junctions of graphene and twisted bilayer graphene \cite{Inbar2023}. Theoretical work has explored the use of the QTM to probe two-dimensional superconductors \cite{Xiao2023}, spin liquids \cite{Peri2024}, as well as spin-ordered states close to metal-insulator transitions \cite{Pichler2024}. 

Beyond electronic dispersions, the QTM is also exquisitely suited to access momentum-resolved dispersions of the collective excitations of van-der-Waals systems as shown by a very recent experiment \cite{Birkbeck2024}.  Whenever the electronic dispersions of tip and sample with their relative twist do not intersect, the momentum mismatch can be bridged by emission of a collective-excitation quantum. Here, we illustrate this modality of the QTM by developing a comprehensive theory of phonon spectroscopy. Our considerations focus on graphene-graphene junctions, but the approach readily generalizes to other junctions and other collective excitations. Along the way, we include analytical results for elastic tunneling between tip and sample for context and comparison. 

Due to the semimetallic nature of graphene, its Fermi circles are typically small for relevant gate-induced densities. Thus,  inelastic tunneling processes enabled by phonon emission dominate except at the smallest twist angles (i.e., for $\theta\gtrsim 5^\circ$ in current experiments \cite{Birkbeck2024}), and the phonon wavevector is approximately equal to the distance between Dirac points of tip and sample. This implies that the relevant phonon wavevector can be systematically varied by changing the twist angle, allowing for direct measurements of the phonon dispersions. 

\begin{figure}[b]
    \centering    \includegraphics[width=.8\linewidth]{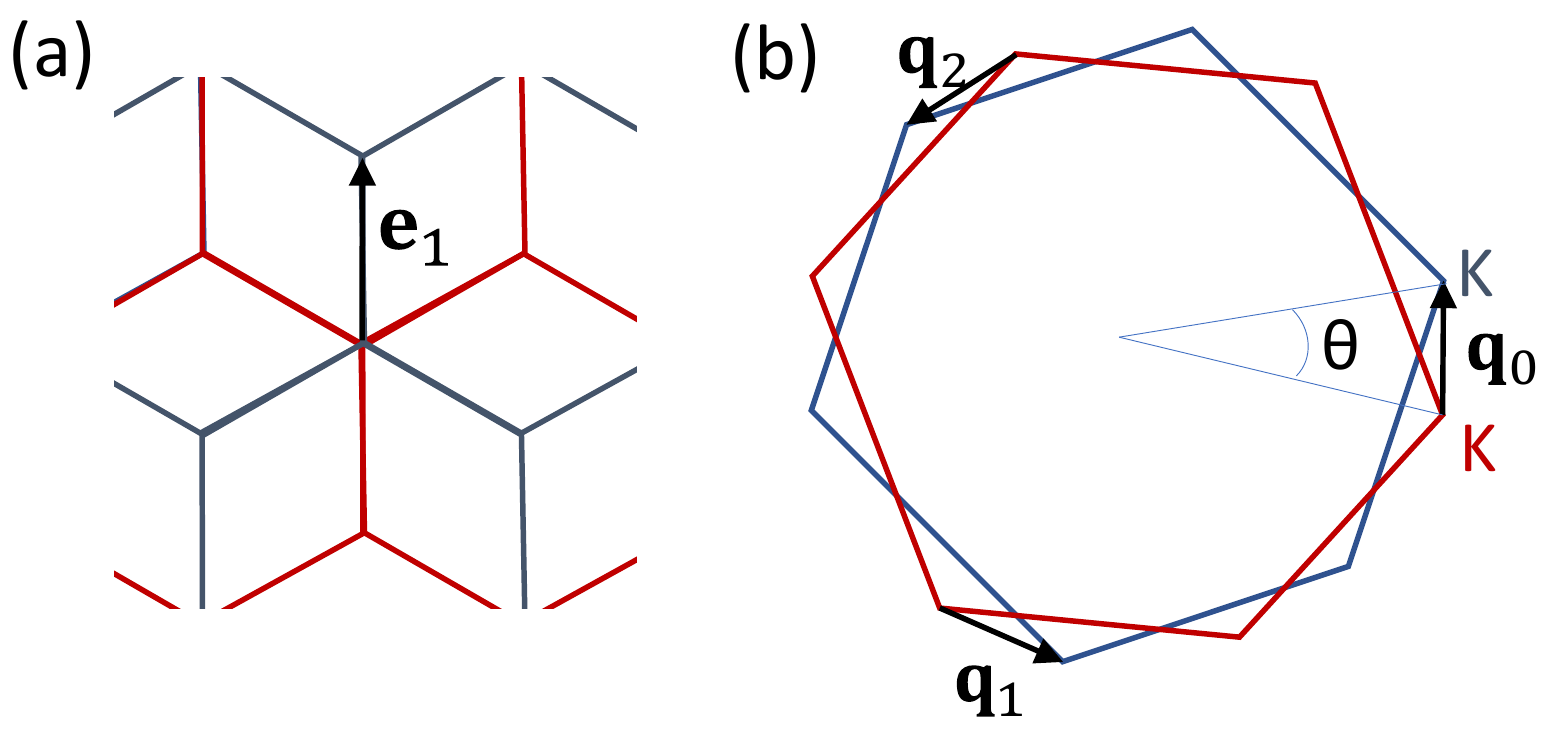} \\
    \includegraphics[width=.8\linewidth]{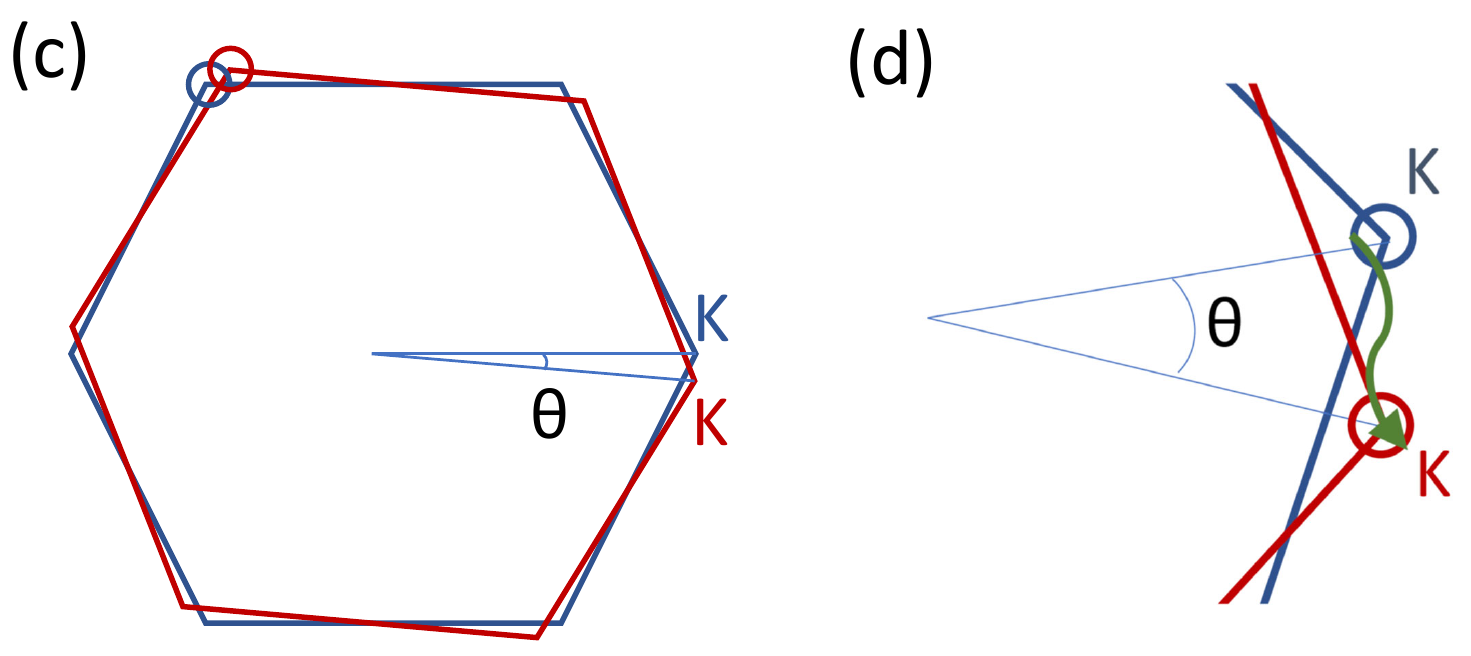}\caption{ (a) Bernal-stacked configuration (two layers shifted relative to each other by the bond vector $\mathbf{e}_1$) used as  reference configuration for twisted layers. Top layer (tip, blue); bottom layer (sample, red). (b) Definition of vectors $\mathbf{q}_j$ connecting the $K$-points of the Brillouin zones of tip and sample. (c) Overlapping Fermi circles of tip and sample at small twist angle $\theta$. In this limit, current is dominated by elastic tunneling between tip and sample. (d) Tunneling at twist angles with mismatched Fermi circles occurs by phonon emission (green arrow connecting states on Fermi circles of tip and sample due to momentum conservation). The phonon wavevector $\mathbf{Q}$ is approximately equal to one of the $\mathbf{q}_j$.}    \label{fig:smallthetaBZ}
\end{figure}

In addition to the phonon dispersion, the measurements also encode the strength of electron-phonon coupling as a function of wavevector. The phonon spectrum and electron-phonon coupling are important inputs in developing a theory of superconductivity  \cite{WMM18,CC18,LBWB19,WHS19,ATF19,CG21,LCR21,CWSS22} and the linear temperature dependence of the resistivity \cite{WHS19,Setal21,IFGL21,SW22,HM22,IL22,AKH23,DCWS23,OF23} in magic-angle twisted bilayer graphene. In both cases, there has been substantial debate whether the phenomenon has a conventional origin in the electron-phonon coupling or an underlying exotic mechanism due to electron-electron correlations. Measurements of graphene-graphene junctions in a QTM may be highly relevant in this context as van-der-Waals coupling tends to lock the interlayer distance  between tip and sample layers of a QTM to the interlayer distance of a twisted bilayer. At the same time, QTM measurements cannot access inelastic phonon processes at very small twist angles, where they will be difficult to differentiate from the elastic-tunneling background. Gleaning information on phonon dispersions and electron-phonon couplings in magic-angle twisted bilayer graphene will thus require extrapolation from larger twist angles. 

This paper is organized as follows. In Sec.\ \ref{sec:overview}, we begin by summarizing the results of the detailed calculations in the subsequent sections. This summary section makes our results accessible without the need to read the more technical sections in detail. Section  \ref{sec:electronic} collects background material and fixes notation. We first introduce the electronic properties of graphene layers (Sec.\ \ref{sec:elgraph}), review tunneling between twisted layers in Secs.\ \ref{sec:tunneling} and \ref{sec:tunneling2}, and discuss the electrostatics of QTM junctions \ref{sec:elst}. Section \ref{sec:eltun} focuses on the elastic tunneling current, giving analytical results for the threshold behaviors of the differential conductance. Section \ref{sec:phonon} discusses the phonon modes and the electron-phonon coupling, including both intralayer and interlayer coupling. The theory of phonon spectroscopy is finally addressed in Sec.\ \ref{sec:inelastic}. Following the general expressions for the inelastic tunneling current in Sec.\ \ref{sec:inelastic1}, we discuss the contributions of the inter- and intralayer electron-phonon coupling in Secs.\ \ref{sec:inelastic2} and \ref{sec:inelastic3}, respectively. We conclude in Sec.\ \ref{sec:conclusions}.

\section{Overview of results}
\label{sec:overview}

\subsection{Tunneling}

We illustrate phonon spectroscopy by considering tunneling between two graphene layers. Twisting the tip and sample layers with respect to each other leads to a relative rotation of their Brillouin zones (Fig.\ \ref{fig:smallthetaBZ}). In particular, this induces a relative displacement of their $K$-points by equal-length vectors $\mathbf{q}_j$, where $j=0,1,2$ enumerates the three equivalent $K$-points within the Brillouin zone [Fig.\ \ref{fig:smallthetaBZ}(b)]. Current flow between tip and sample is dominated by elastic tunneling as long as the Fermi circles of tip and sample intersect \cite{Inbar2023} [Fig.\ \ref{fig:smallthetaBZ}(c)]. This is a consequence of the fact that tunneling between tip and sample conserves crystal momentum modulo reciprocal lattice vectors of the graphene layers. At larger twist angles, the momentum mismatch between the electrons in tip and sample requires inelastic processes involving phonon emission, with the phonon wavevector approximately equal to one of the $\mathbf{q}_j$ [Fig.\ \ref{fig:smallthetaBZ}(d)]. (Here, we assume that temperature is sufficiently low that phonon absorption can be neglected.) Bias voltages, at which $eV_b$ equals the energy $\hbar\omega_{r,\mathbf{q}_j}$ of a phonon mode $r$ are associated with threshold features in the current-voltage characteristic. Tracking these inelastic-tunneling features as a function of twist angle $\theta$ (and hence $\mathbf{q}_j$) allows for mapping out the phonon dispersion along certain lines in momentum space. (Umklapp processes lead to additional sharp elastic scattering peaks at larger, commensurate twist angles due to overlap of Fermi surfaces in higher Brillouin zones \cite{Inbar2023}.) 

\begin{figure}[b]
    \centering    \includegraphics[width=.8\linewidth]{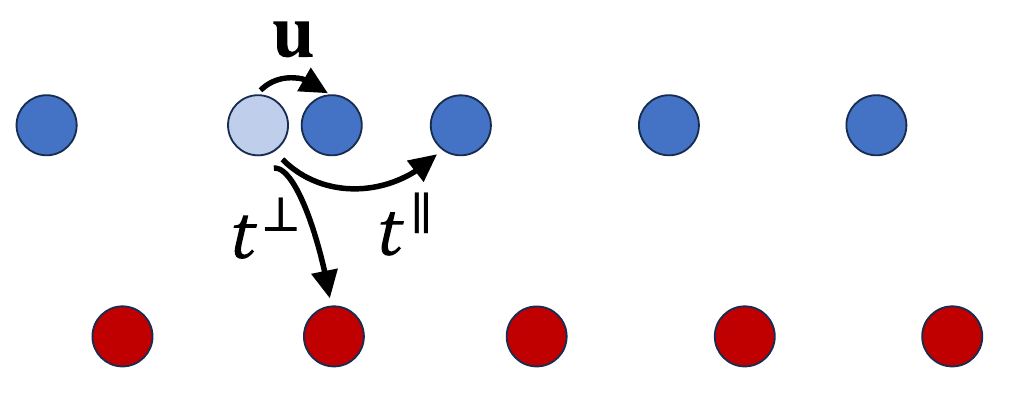}\caption{Electron-phonon coupling in twisted graphene layers. Sketch of tip (blue) and sample (red) layers with phonon-induced atomic displacement $\mathbf{u}$ (illustrated for the in-plane displacement of one atom in the tip). The displacements $\mathbf{u}$ modify the bond lengths and consequently the hopping amplitudes $t^\parallel$ within the layer (intralayer electron-phonon coupling $\mathcal{H}_\mathrm{intra}$) and  between layers $t^\perp$ (interlayer electron-phonon coupling $\mathcal{H}_\mathrm{inter}$). }
    \label{fig:ephcoupl}
\end{figure}

Electron-phonon coupling emerges from several mechanisms \cite{SA02}. 
Modifications of the hybridization of carbon orbitals associated with phonon-induced changes in the bond lengths contribute as illustrated in Fig.\ \ref{fig:ephcoupl}. 
The intralayer coupling $\mathcal{H}_\mathrm{intra}$ originates in changes of the hopping amplitudes within the layer, corresponding to electron-phonon coupling within the individual graphene layers. In twisted bilayers, phonons also affect the amplitude of interlayer tunneling, giving rise to the interlayer electron-phonon coupling $\mathcal{H}_\mathrm{inter}$. 
In addition, longitudinal acoustic phonons lead to a local expansion or contraction of the lattice, which shifts the chemical potential \cite{SA02,NGPNG09}, referred to as defomation potential. We find that the coupling mechanism dominating the tunneling current differs between phonon modes, so that it is important to account for the various couplings to understand phonon signatures in QTM measurements. 

Twisted graphene layers are described by the Hamiltonian
\begin{equation}
\mathcal{H}=\mathcal{H}_0 + \mathcal{H
    }_T + \mathcal{H
    }_\mathrm{intra} + \mathcal{H
    }_\mathrm{inter}. 
\end{equation}
Here, $\mathcal{H}_0$
describes the two uncoupled graphene layers, including their phonon modes, and
$\mathcal{H}_T$ accounts for the (purely electronic) interlayer tunneling. Retaining terms to first order in the interlayer tunneling, we can then expand the $\mathcal{T}$-matrix for electron scattering between tip and sample layers as
\begin{equation}
    \mathcal{T} = \mathcal{H}_T + \mathcal{H}_\mathrm{inter} + \mathcal{H}_T\mathcal{G}_0\mathcal{H}_\mathrm{intra}+\mathcal{H}_\mathrm{intra}\mathcal{G}_0\mathcal{H}_T +\ldots 
    \label{eq:Tmatrix}
\end{equation}
The first term $\mathcal{H}_T$ on the right hand side gives rise to the elastic tunneling processes at small twist angles. Inelastic tunneling involving the emission of a phonon due to the interlayer electron-phonon coupling is described by the second term. Both of these processes can be described in the lowest order in a Fermi-golden-rule calculation of the tunneling current. The remaining two terms describe higher-order inelastic processes involving both electron tunneling and intralayer electron-phonon coupling, with the Green function $\mathcal{G}_0=[E-\mathcal{H}_0]^{-1}$ of the uncoupled layers accounting for the energy denominators of the virtual intermediate states. We find that the inelastic tunneling current can be dominated by one electron-phonon coupling or the other, despite their different orders in perturbation theory. 

\subsection{Electrostatics and characteristic voltages}

Due to the small quantum capacitance of the graphene layers, a bias voltage applied between tip and sample will predominantly modify the chemical potentials. This is accompanied by a smaller relative shift $e\phi$ in energy of the Dirac points of tip and sample due to the electrostatic potential difference $\phi$. The ratio  of the shifts in electrostatic and chemical potentials is of order $q_\mathrm{TF}d$, where $q_\mathrm{TF}$ is the Thomas-Fermi screening wavevector of graphene and $d$ the distance between tip and sample layers [see Eq.\ (\ref{eq:phimuTF}) below and the discussion around it for more details]. In our analytical calculations, we focus on the limit in which  $q_\mathrm{TF}d\ll 1$ (small quantum capacitance) and assume overall charge neutrality. Then, tip and sample have opposite chemical potentials $\pm\mu$, so that the bias voltage (i.e., the difference in electrochemical potential) is $eV_b=2\mu+e\phi$ with $e\phi\ll 2\mu$.

\begin{figure}[b]
    \centering
    \includegraphics[width=.9\linewidth]{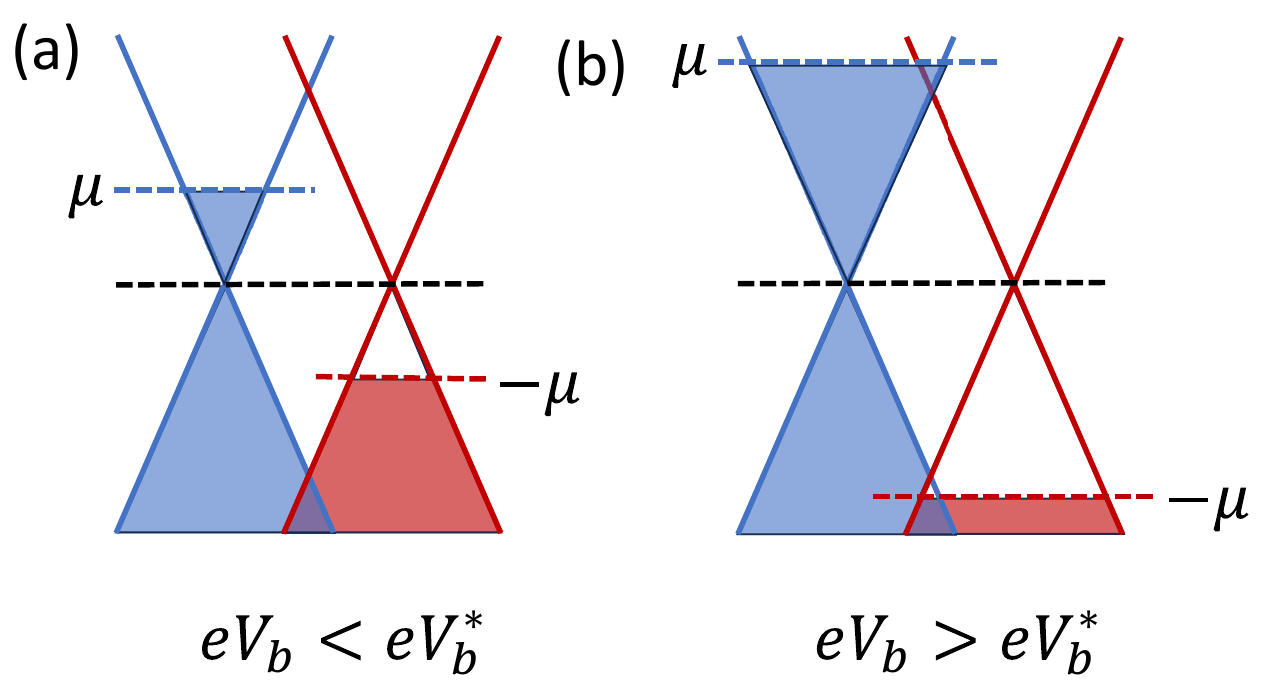} \includegraphics[width=.85\linewidth]{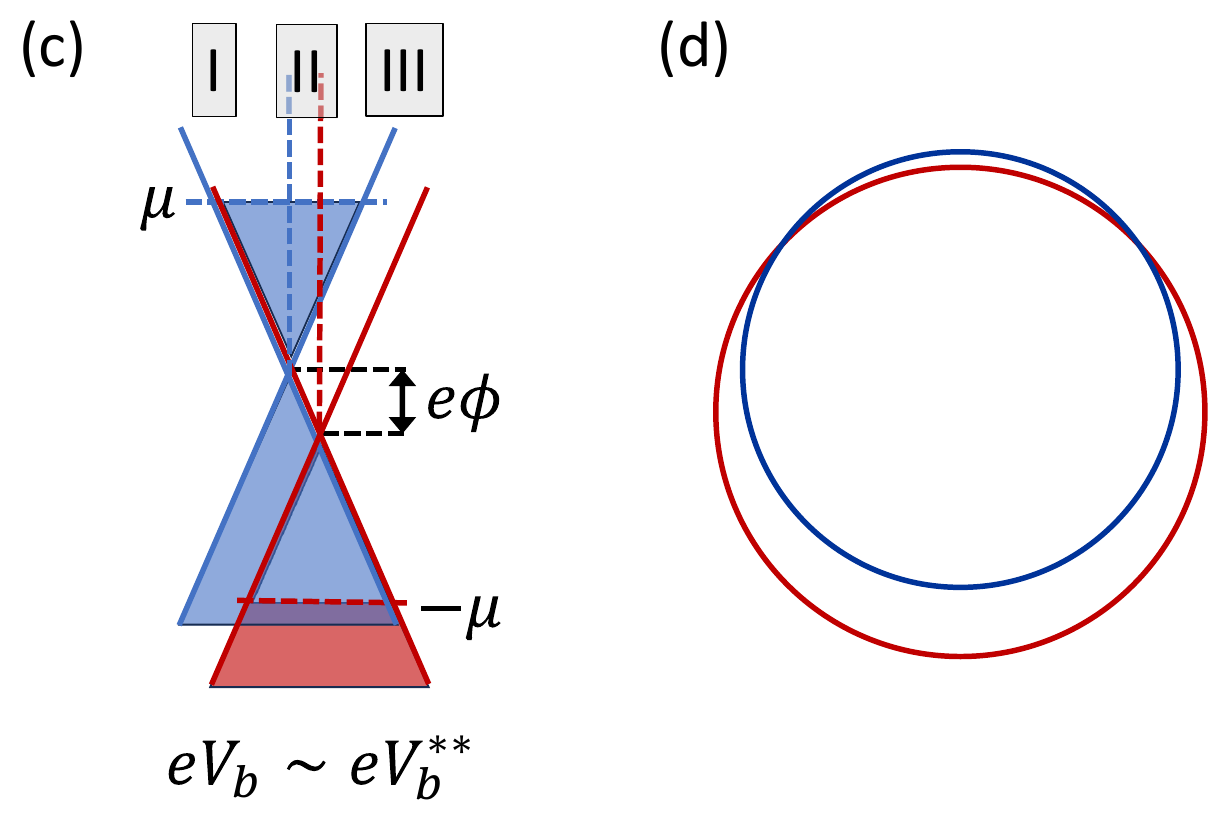}
    \caption{Characteristic voltages for elastic QTM tunneling at small twist angles, assuming overall charge neutrality of tip and sample as well as small quantum capacitance. (a,b) Tunneling onsets at $eV_b^*$ [Eq.\ (\ref{eq:vstar})], beyond which occcupied states in the Dirac cone of tip overlap with empty states in the Dirac cone of the sample. Illustration of Dirac cones of tip and sample separated in momentum by $q_0$ and their fillings for a voltage (a) below and (b) above $V_b^*$. (c) Dirac cones of tip and sample for bias voltages of order $V_b^{**}$ exhibiting nesting due to the relative electrostatic shift $e\phi$ of the Dirac points. I, II, III refer to momentum regions to the left of the blue dashed line (I), between blue and red dashed line (II), and to the right of the red dashed line (III). (d) Constant-energy cut through the shifted Dirac-cone dispersions of tip and sample for bias voltages close to the nesting condition illustrated in panel (c).}
    \label{fig:smalltheta}
\end{figure}

At small bias voltages, the electrostatic shift $\phi$ can be neglected and the Dirac points of tip and sample are aligned in energy, but offset in momentum by $\mathbf{q}_j$.  The offset Dirac cones intersect at energies that are larger in magnitude than $\hbar v_D q_0/2$, where $v_D$ is the Dirac velocity. As the chemical potentials of tip and sample are equal to $\pm eV_b/2$, this leads to a characteristic voltage of 
\begin{equation}
  eV_b^*=\hbar v_Dq_0   
\end{equation}
for elastic scattering [Fig.\ \ref{fig:smalltheta}(a) and (b)], with current flow due to elastic tunneling only possible for bias voltages $V_b>V_b^*$. 

At larger bias voltages, the electrostatic potential leads to an appreciable relative shift of the Dirac points in energy. When this shift $e\phi$ becomes of order $\hbar v_D q_j$, there is approximate nesting of the Dirac dispersions of tip and sample [Fig.\ \ref{fig:smalltheta}(c)]. Nesting defines a second characteristic voltage $V_b^{\ast\ast}$ through the condition
\begin{equation}
e\phi(V_b^{\ast\ast})=\hbar v_D q_0.
\end{equation}
Note that there is a sharp drop in the elastic tunneling current as the bias increases past this characteristic voltage. 

The dispersions depicted in Fig.\ \ref{fig:smalltheta}(a) and (b) neglect interlayer tunneling, which opens gaps at the crossing points of the Dirac dispersions of the two layers. The magnitude of the resulting gaps is given by the interlayer tunneling strength $w$. Our perturbative approach requires that these gaps be small compared to $eV_b^*$. This is satisfied for twist angles 
\begin{equation}
  \theta > \frac{w}{v_D|\mathbf{K}|}.   
  \label{eq:twistangle}
\end{equation}
(Here, $\mathbf{K}$ denotes the location of the $K$-point as measured from the $\Gamma$-point.) This is equivalent to the condition that the twist angle be larger than the magic angle of twisted bilayer graphene. 

For twist angles satisfying Eq.\ (\ref{eq:twistangle}), inelastic tunneling processes are relevant provided that the phonon energy $\hbar\omega_{r,\mathbf{q}_j}$ is smaller than $eV_b^*$. In this case, the tunneling current at bias voltages $V_b<V_b^*$ is entirely due to inelastic processes, allowing for  measurements of the phonon dispersions and the electron-phonon coupling. For optical phonons with frequencies   $\sim 200$meV, this gives a minimal twist angle of 1-2$^\circ$. 
For acoustic phonons, the condition is less stringent due to their smaller energy. There exists an energy window as long as the Dirac velocity is large compared to the mode velocity of the acoustic phonons, which is always the case away from the magic angle.  

\subsection{Elastic tunneling}

Coherent tunneling across an extended contact (area $\Omega = L_x L_y$) is central to the QTM \cite{Inbar2023}.  To bring out the importance of coherence, we consider a reference problem of $\Omega/\lambda_F^2$ parallel incoherent local tunneling contacts. The Fermi wavelength $\lambda_F=2\pi/\kappa_F$ defines the characteristic size of the local contacts, so that the coarse-grained local tunneling of a contact at $\mathbf{R}$ can be approximated as $w\lambda_F^2\delta(\mathbf{r}-\mathbf{R})$. Here, $w$ denotes the underlying tunneling amplitude of the extended contact with units of energy. This gives an incoherent  tunneling current of order \begin{equation}
    (2\pi e/\hbar)(w\lambda_F^2)^2\nu(\mu)[N_f\nu(\mu)eV_b]
\end{equation} 
per tunnel contact, where $\nu(\mu) = \mu/(2\pi\hbar ^2 v_D^2)$ is the density of states per flavor at the chemical potential. This expression for the current is composed of the Fermi-golden-rule rate for tunneling of an incident electron and the number of incident electrons within the voltage window accounting for the spin and valley degeneracy $N_f=4$. With the linear graphene dispersion $E=\hbar v_D \kappa$ and the relation $\mu=\hbar v_D \kappa_F = eV_b/2$ between voltage and chemical potential (for sufficiently small voltages, so that the electrostatic potential $\phi$ can be neglected), this gives a differential-conductance scale of order
\begin{equation}
    G_\mathrm{incoh}= \frac{e^2}{h} \frac{N_f w^2\Omega}{\hbar^2v_D^2}
\end{equation}
for an array of 
$\Omega/\lambda_F^2$ parallel contacts. 

We find that $G_\mathrm{incoh}$ provides a convenient scale for expressing our results for the tunneling current in the QTM. Elastic tunneling sets in at voltages $V_b>V_b^*$. The differential conductance in the vicinity of $V_b^*$ is given by (see Sec.\ \ref{sec:Vast})
\begin{equation}
    \frac{dI}{dV_b}= 3\sqrt{2} G_\mathrm{incoh} \sqrt{\frac{V_b^*}{V_b-V_b^*}} \theta\!\left(V_b -V_b^* \right),
    \label{eq:finvast}
\end{equation}
exhibiting a square-root divergence in the bias voltage. As $eV_b^\ast = \hbar v_D q_0$, measuring the threshold voltage as a function of twist angle (and hence $\mathbf{q}_0$) can be viewed as direct spectroscopy of the Dirac dispersion of the graphene layers. 

Similarly, the current near the threshold voltage $V_b^{\ast\ast}$ for nesting is 
(see Sec.\ \ref{sec:Vastast})
\begin{equation}
    I = \frac{3 }{2}
G_\mathrm{incoh}\frac{(V_b^{**}) ^2
    }{\sqrt{V_b^*|\phi(V_b^{**})-\phi(V_b)|}} \theta(V_b^{**} -V_b). 
    \label{eq:finvastast}
\end{equation}
This implies a large negative differential conductance at voltage $V_b^{\ast\ast}$. The divergence at $V_b^{\ast\ast}$ is a consequence of the assumptions of strictly linear dispersion and momentum-conserving tunneling. The tunneling current originates from momenta, where the Dirac cones of tip and sample touch, so that the contributions to the current diverge concurrently at all energies within the bias window [Fig.\ \ref{fig:smalltheta}(c,d)]. The divergence of the current at $V_b^{**}$ will thus be cut off by nonlinear corrections to the dispersion relation as well as spatial inhomogeneities, which lift strict momentum conservation.

\begin{table*}[]
\setlength{\tabcolsep}{5pt}
    \centering
    \begin{tabular}{||c|c|c|c|c|c||}
    \hline
     $dI/dV$ & \multicolumn{2}{|c|}{acoustic} & \multicolumn{2}{|c|}{optical} & \multicolumn{1}{|c||}{out-of-plane} 
     \\
        $\left[G_\mathrm{incoh}\theta(eV_b-\hbar\omega_{\mathbf{q}_0})\right]$ & LA & TA & LO & TO & ZO$^\prime$ \& ZO \\
        \hline\hline
        $\mathcal{H}_\mathrm{inter}$ & $(\kappa_F\ell_\mathrm{ZPM})^2 \sin^2\theta$ & $(\kappa_F\ell_\mathrm{ZPM})^2 \cos^2\theta$ & $(\kappa_F\ell_\mathrm{ZPM})^2\sin^2\theta$ & $(\kappa_F\ell_\mathrm{ZPM})^2\cos^2\theta$ & $(\kappa_Fa)^2 (\ell_\mathrm{ZMP} \frac{\partial\ln w}{\partial d})^2$  \\
        Eq.\ (\ref{eq:ResultInter}) &  &  &  &  &  \\ 
\hline
        $\mathcal{H}_\mathrm{intra}$ & $(\kappa_F\ell_\mathrm{ZPM})^2$ & $(\kappa_F\ell_\mathrm{ZPM})^2$ & $\left(\frac{\kappa_F\ell_\mathrm{ZPM}}{q_0 a}\right)^2$ & $\left(\frac{\kappa_F\ell_\mathrm{ZPM}}{q_0 a}\right)^2$ & higher order \\
              Eq.\ (\ref{eq:result_intra})  &  &  &  &  &    \\ 
        \hline
    \end{tabular}
    \caption{Summary of results for the step in $dI/dV$ at the inelastic threshold $eV_b=\hbar\omega_{\mathbf{q}_0}$ for various phonon modes. Expressions given in the table give the parametric dependences when multiplied by $G_\mathrm{incoh}\theta(eV_b-\hbar\omega_{\mathbf{q}_0})$. 
    The contribution of $\mathcal{H}_\mathrm{inter}$ is obtained in first-order perturbation theory, while the contribution of $\mathcal{H}_\mathrm{intra}$ follows from a second-order calculation treating both $\mathcal{H}_\mathrm{intra}$ and $\mathcal{H}_T$ perturbatively, see Eq.\ (\ref{eq:Tmatrix}). The equation numbers given in the table point to the full expressions in the text. We include the results for electron-phonon coupling originating from modifications of hopping amplitudes by phonons. For LA phonons, there is an additional contribution  due to the deformation potential (strength $D$), which can be obtained from the contribution of $\mathcal{H}_\mathrm{intra}$ by the replacement  
    $\partial t^\parallel/{\partial \ln a}  \leftrightarrow D$, see Sec.\ \ref{sec:deformation}. We note that the amplitude $\ell_\mathrm{ZPM}$ of the zero-point motion contains an implicit dependence on the  frequency of the phonon mode under consideration, cp.\ Eq. (\ref{eq:ZPM_intro}).}
    \label{tab:elph}
\end{table*}

\subsection{Inelastic tunneling}
\label{sec:qual_inelastic}

When the Fermi circles of tip and sample no longer intersect at larger twist angles, momentum-conserving tunneling at low bias voltages is enabled by emission (or absorption) of phonons. 
Neglecting the small electronic momenta relative to the $K$-points, the relevant phonons have wavevectors equal to the wavevectors $\mathbf{q}_j$ connecting the $K$-points of the tip and the sample [Fig.\ \ref{fig:smallthetaBZ}(b)]. Consequently, the inelastic tunneling channels for the various phonon modes $r$ open beyond the threshold voltages $eV_b=\hbar\omega_{r,\mathbf{q}_j}$, where $\omega_{r,\mathbf{q}_j}$ denotes the phonon frequency of mode $r$ at wavevector $\mathbf{q}_j$. This appears in the tunneling current as steps in the differential conductance and consequently as peaks in $d^2I/dV_b^2$. 

For out-of-plane (flexural) phonons, the electron-phonon coupling is dominated by the  interlayer electron-phonon coupling $\mathcal{H}_\mathrm{inter}$. While  $\mathcal{H}_\mathrm{inter}$ depends linearly on the phonon displacements, the intralayer coupling $\mathcal{H}_\mathrm{intra}$ to out-of-plane phonons has a weaker quadratic dependence on the mode displacements. The dominant coupling due to $\mathcal{H}_\mathrm{inter}$ 
emerges directly from the change in the interlayer tunneling amplitude due to the out-of-plane component of the atomic displacements and is thus of order $\ell_\mathrm{ZPM}(\partial w/\partial d)$. Here, $d$ is the interlayer distance and   \footnote{Notice that the actual real-space amplitude of the zero-point fluctuations for a particular phonon mode is smaller than $\ell_\mathrm{ZPM}$ by a factor $1/\sqrt{N}$.}
\begin{equation}
    \ell_\mathrm{ZPM}=(\hbar/M\omega_{r,\mathbf{q}_0})^{1/2}
    \label{eq:ZPM_intro}
\end{equation}
(with $M$ the mass per unit cell of a single layer)
the zero-point-motion amplitude of the relevant out-of-plane phonon mode. We then find [Eq.\ (\ref{eq:ResultInter})]
\begin{equation}
     \frac{d^2I}{dV_b^2} \sim G_\mathrm{incoh} (\kappa_F a)^2
     \left( \ell_\mathrm{ZPM}\frac{\partial \ln w}{\partial d}\right)^2\delta(V_b- \hbar \omega_{r,\mathbf{q}_0}/e),   \label{eq:phonon1st}
\end{equation}
where the factor $(\kappa_F a)^2$ accounts for the size of the graphene Fermi circles. ($a$ is the carbon-carbon bond length of graphene.) 

There are two out-of-plane phonon modes $r$, which we denote by ZO$^\prime$ and ZO, respectively. The mode ZO$^\prime$ is an optical mode in the out-of-plane direction, i.e., the atomic displacements are antisymmetric between the two layers. At the same time, it is acoustic in nature within the plane, i.e., the long-wavelength displacements of the two sublattices are symmetric within each layer. Even when mechanical coupling between the tip and sample layers is negligible, the mode frequencies saturate to a constant at small wavevectors due to mechanical coupling between the graphene layers and their substrates. At larger wavevectors, the mode frequencies are quadratic in the phonon wavevector akin to the flexural modes of free-standing graphene membranes. The ZO mode is also optical in nature within the plane and thus characterized by larger mode frequencies for all wavevectors. 

\begin{figure}[b]
    \centering \includegraphics[width=.8\columnwidth]{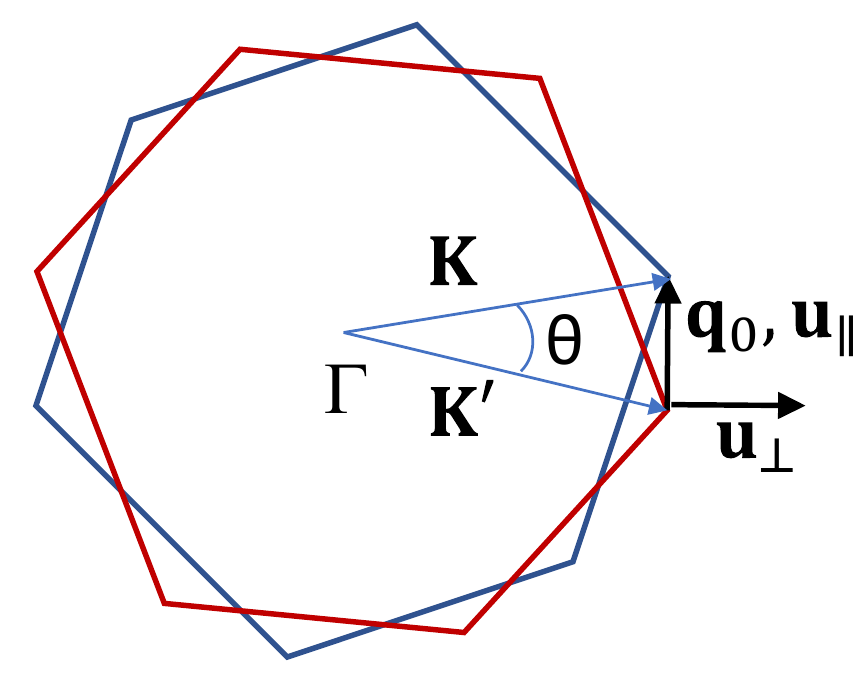}
    \caption{Illustration of the interlayer electron-phonon coupling. At small twist angles $\theta$, the phonon wavevector $\mathbf{q}_0$ is almost perpendicular to the vectors $\mathbf{K}$ and $\mathbf{K}'$ of the $K$-points of tip and substrate, respectively. As a result, the phonon displacements are approximately parallel to $\mathbf{K}$ and $\mathbf{K}'$ for transverse phonons ($\mathbf{u}_\perp$) and perpendicular for longitudinal phonons ($\mathbf{u}_\parallel$). }
    \label{fig:transverse}
\end{figure}

The interlayer electron-phonon coupling to in-plane phonons has a different origin. A relative displacement $\mathbf{u}$ of the two layers does not modify the magnitude of the interlayer tunneling, but changes its phase by $\exp(i\mathbf{K}\cdot\mathbf{u})$ \cite{Bistritzer2011}. If $\mathbf{u}$ originates from a phonon displacement, the phase becomes time dependent, akin to a time-dependent vector potential. Thus, $\mathcal{H}_\mathrm{inter}$ can be viewed as being due to a synthetic electric field \cite{OGM09,O19,MN20}. Importantly, the coupling is maximal for mode displacements $\mathbf{u}$, which are parallel to the vector $\mathbf{K}$ of the $K$-point. At small twist angles, the phonon wavevector $\mathbf{q}_0$ is approximately  perpendicular to $\mathbf{K}$, so that the coupling is predominantly to transverse phonon modes. This is illustrated in Fig.\ \ref{fig:transverse}. Unlike the intralayer electron-phonon coupling, this coupling does not go to zero in the long-wavelength limit. Transverse phonons, both acoustic (TA) and optical (TO), will thus involve an interlayer electron-phonon coupling of order 
$w \mathbf{K}\cdot \mathbf{u} \sim w (\ell_\mathrm{ZPM}/a)\cos\theta$ from expanding the phase factor to linear order. Combining this with the factor accounting for the size of the graphene Fermi circles gives
\begin{equation}
     \left.\frac{d^2I}{dV_b^2}\right|_\mathrm{inter,T} \sim G_\mathrm{incoh} (\kappa_F \ell_\mathrm{ZPM})^2 \cos^2\theta \delta(V_b-
       \hbar\omega_{r,\mathbf{q}_0}/e)   \label{eq:phonon1stT}
\end{equation}
for the contribution of the interlayer electron-phonon coupling for transverse acoustic and optical phonons. 
The expression for longitudinal acoustic (LA) and optical (LO) phonons is similar, differing only in its twist-angle dependence, 
\begin{equation}
     \left.\frac{d^2I}{dV_b^2}\right|_\mathrm{inter, L} \sim G_\mathrm{incoh} (\kappa_F \ell_\mathrm{ZPM})^2 \sin^2\theta \delta(V_b-
       \hbar\omega_{r,\mathbf{q}_0}/e).  \label{eq:phonon1stL}
\end{equation}
The full result for both longitudinal and transverse phonons is given in Eq.\ (\ref{eq:ResultInter}). 

The intralayer electron-phonon coupling $\mathcal{H}_\mathrm{intra}$ gives contributions of the same order for acoustic phonons, albeit with a different twist-angle dependence. We can estimate the relevant electron-phonon coupling by noting that $\mathcal{H}_\mathrm{intra}$ originates in the dependence of $t^\parallel$ on the bond length. Thus, the intralayer electron-phonon coupling is of order $\frac{\partial t^\parallel}{\partial a}\ell_\mathrm{ZPM}(q_0 a)$ for acoustic phonons. Here, the last factor accounts for the fact that the relative displacements of neighboring atoms is suppressed in the long-wavelength limit. Combining this with the fact that the tunneling Hamiltonian is of order $w$ and the energy denominators are of order $\hbar v_D q_0$, we find that the contribution of $\mathcal{H}_\mathrm{intra}$ to the $\mathcal{T}$-matrix is of order 
\begin{equation}
    \frac{w (\partial t^\parallel/\partial a)\ell_\mathrm{ZPM}(q_0 a)}{\hbar v_D q_0}. 
\end{equation}
With the estimates $\partial t^\parallel/\partial a \sim t^\parallel/a$ and $\hbar v_D/a \sim t^\parallel$, this becomes of order $w(\ell_\mathrm{ZPM}/a)$, which is indeed of the same order as the contribution of $\mathcal{H}_\mathrm{inter}$. At the same time, $\mathcal{H}_\mathrm{intra}$ has only weak twist-angle dependence and is of the same order for longitudinal and transverse phonons due to the triad of bond vectors. This yields
\begin{equation}
     \left.\frac{d^2I}{dV_b^2}\right|_\mathrm{intra} \sim G_\mathrm{incoh} (\kappa_F \ell_\mathrm{ZPM})^2 \delta(V_b-
       \hbar\omega_{r,\mathbf{q}_0}/e).  \label{eq:phonon2A}
\end{equation}
for acoustic phonons. 

For optical phonons, the intralayer electron-phonon coupling $\mathcal{H}_\mathrm{intra}$ does not involve the suppression factor $q_0 a$, so that 
\begin{equation}
     \left.\frac{d^2I}{dV_b^2}\right|_\mathrm{intra} \sim G_\mathrm{incoh} \left(\frac{\kappa_F \ell_\mathrm{ZPM}}{q_0 a}\right)^2 \delta(V_b-
       \hbar\omega_{r,\mathbf{q}_0}/e).  \label{eq:phonon2O}
\end{equation}
At long wavelengths, this dominates over the contribution of $\mathcal{H}_\mathrm{inter}$ in Eqs.\ (\ref{eq:phonon1stT}) and (\ref{eq:phonon1stL}). The full result for the contribution of $\mathcal{H}_\mathrm{intra}$ for both acoustic and optical phonons is given in Eq.\ (\ref{eq:result_intra}).

We summarize these results for the inelastic contributions to the tunneling current in Table \ref{tab:elph}. We conclude this section by addressing the limit of small twist angles $\theta$ (but sufficiently large that elastic scattering can be neglected). For acoustic modes, $q_0 \sim \theta$ and $\omega_{r,\mathbf{q}_0} \sim q_0$, which implies $\ell_\mathrm{ZPM} \sim 1/\sqrt{q_0}$. For overall charge neutrality, the threshold condition $eV_b = \hbar \omega_{q_0}$ implies that $\kappa_F \sim q_0$, while $\kappa_F$ approaches a nonzero constant for small $q_0$ away from overall charge neutrality. We thus find that for acoustic modes, $d^2I/dV_b^2$ diverges as $1/q_0$ at small twist angles away from charge neutrality. For optical modes in the limit $q_0\to 0$, $\kappa_F$ and $\ell_\mathrm{ZPM}$ approach nonzero constants, so that one finds a $1/q^2_0$ divergence both at and away from charge neutrality. The divergence is cut off by the condition for the twist angle in Eq.\ (\ref{eq:twistangle}), which corresponds to $q_0\gtrsim w/v_D$. A further limitation specific to optical phonons is set by Eq.\ (\ref{eq:twistoptlim}).

\begin{figure}[t]
    \centering
    \includegraphics[width=.9\columnwidth]{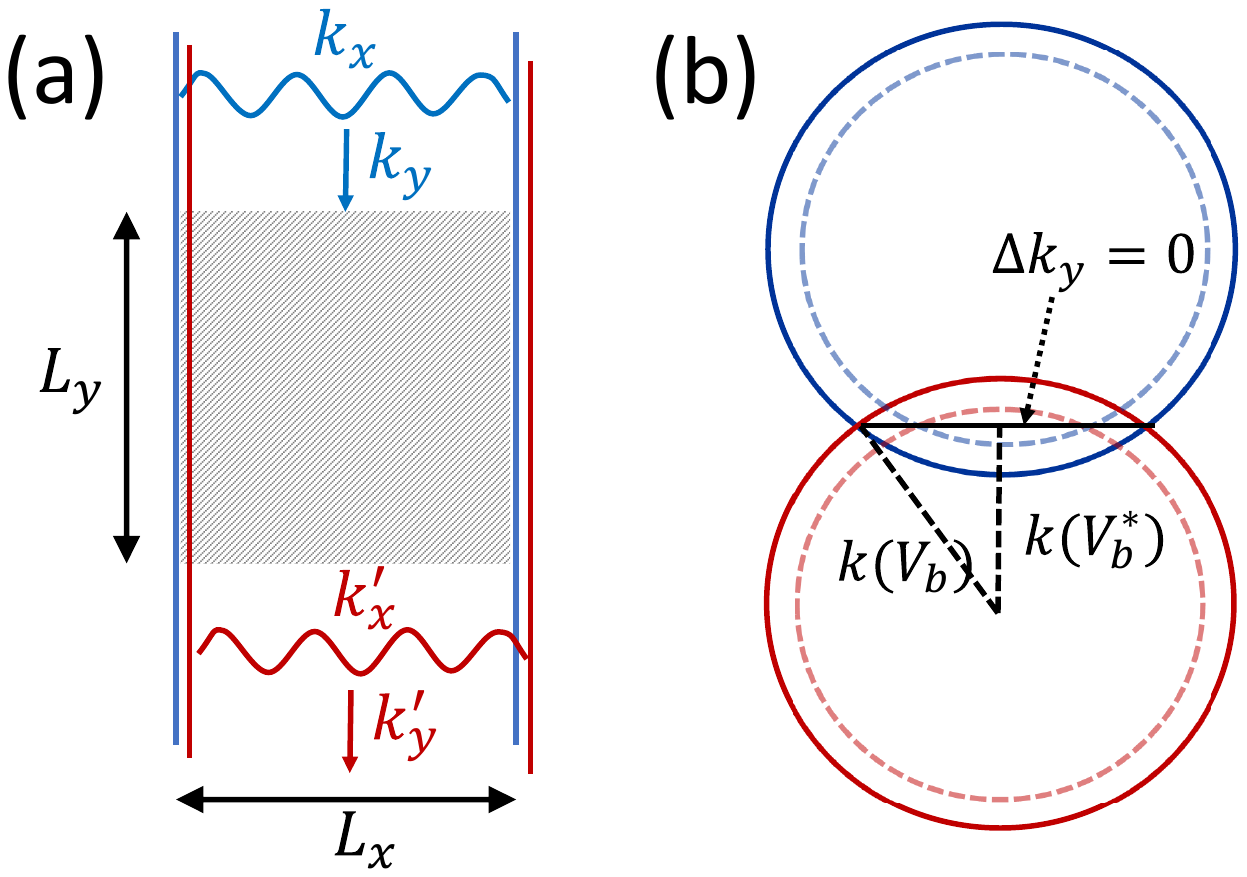}
    \caption{(a) Scattering geometry of the QTM junction between tip (blue) and sample (red). Plane waves in the tip layer with wavevector $k_y$ in channel $k_x$ impinge on the tunneling contact (black shaded area) and are transmitted into plane waves with momentum $k_y^\prime$ in channel $k_x^\prime$ in the sample. (b) Constant-energy lines of the dispersions of tip and sample for bias voltage $V_b$ close to the characteristic bias voltage $V_b^*$ (full colored lines: energy $E=eV_b$; dashed colored lines: $eV_b>E>eV_b^*$).  
    Tunneling occurs at the crossing points of the dispersions of tip (blue) and sample (red) along the line $\Delta k_y=0$ (black). The length $2k_\mathrm{max}$ of this line at a given bias voltage $V_b$ can be obtained from the indicated right triangle.}
    \label{fig:scatteringgeometry}
\end{figure}

\subsection{Scattering picture}
\label{sec:overview_scat}

The current can be estimated using a scattering approach, which gives insight beyond the Fermi-golden-rule calculations presented in subsequent sections. We consider a scattering geometry, in which electrons in the tip impinge on the tip-sample junction in the $y$-direction, with channels defined by wavevectors $k_x$ (Fig.\ \ref{fig:scatteringgeometry}). When normalizing the incoming and outgoing states to unit flux in the $y$-direction, the tunneling current is 
\begin{equation}
 I = \frac{e}{h}\sum_{k_xk_x'}\int dE dE' T_{k_xk_x'}(E,E')f_\mu(E)[1-f_{-\mu}(E')], 
 \label{eq:currentstart}
\end{equation}
where we assume $V_b>0$ and zero temperature. Here, $T_{k_xk_x'}(E,E')$ is the probability density that an electron in channel $k_x$ of the tip impinging on the junction at energy $E$ is scattered into channel $k_x'$ of the sample at energy $E'$. 

For elastic tunneling, we can approximate 
\begin{equation}
T_{k_x k_x^\prime}(E,E') \sim \frac{w^2}{\hbar^2|v_y v_y^\prime|}
\left|\int_\Omega d\mathbf{r} \frac{e^{i\mathbf{k}\cdot\mathbf{r}-i\mathbf{k}'\cdot\mathbf{r}}}{L_x}\right|^2 \delta(E-E')
\end{equation}
to lowest order in the tunneling amplitude (Born approximation). Here, we defined the mode velocities $v_y=(1/\hbar)(\partial E/\partial k_y)$ and 
$v_y^\prime = (1/\hbar)
(\partial E'/\partial k_y^\prime)$ of tip and sample, respectively. Focusing on order-of-magnitude estimates for small twist angles, we have suppressed the sublattice structure of the graphene wave functions and consider only momentum-conserving tunneling at the $K$-point, i.e., we neglect umklapp processes. Evaluating the integral and taking the limits of large $L_x$ and $L_y$ gives
\begin{equation}
T_{k_x k_x^\prime}(E,E') \sim \frac{w^2}{\hbar^2|v_y v_y^\prime|}
\delta_{k_x,k_x^\prime} L_y \delta(k_y-k_y^\prime)
\delta(E-E'),
\end{equation}
which makes the momentum-conserving nature of tunneling explicit. Here, $k_y=k_y(k_x,E)$ and $k_y^\prime=k_y^\prime(k_x^\prime,E')$ are determined by the electron dispersions in tip and sample, respectively.  Inserting $T_{k_x k_x^\prime}(E,E')$ into the expression for the current and performing the energy integrals, we find
\begin{equation}
  I \sim \int_{\Delta k_y=0} dk_x \frac{ew^2\Omega}{\hbar^2 |v_y -v_y^\prime|}\theta(k_\mathrm{max}-|k_x|).
  \label{eq:currint}
\end{equation}
For convenience, we temporarily measure $k_x$ from the line connecting the Dirac points of tip and sample. The integral is over the line defined by $\Delta k_y = k_y - k_y^\prime = 0$ and we used that $\partial \Delta k_y/\partial E = 1/\hbar v_y - 1/\hbar v_y^\prime$. 

For voltages close to $V_b^\ast$, the dispersions of tip and sample are illustrated in Fig.\ \ref{fig:scatteringgeometry}(b). The condition $\Delta k_y = 0$ enforces $k_y=k_y^\prime = 0$, so that $v_y^\prime = -v_y \simeq v_D$. Moreover, one reads off $\hbar v_D k_\mathrm{max} = [(eV_b/2)^2-(eV_b^\ast/2)^2]^{1/2}$.
Inserting these relations into Eq.\ (\ref{eq:currint}), we recover the parametric dependences in Eq.\ (\ref{eq:finvast}).

For bias voltages near $V_b^{**}$, the dispersions of tip and sample are illustrated in Fig.\ \ref{fig:smalltheta}(d). Approximate nesting of the Fermi circles of tip and sample implies that $v_y$ is close to  $v_y^\prime$, leading to a strongly enhanced current. Nesting occurs for any energy $E$ within the bias window, with the current dominated by energies  $|E| > e\phi (V_b^{\ast\ast})$. For these energies, one readily estimates
\begin{equation}
    \frac{|v_y-v_y^\prime|}{v_D}\sim \left(\frac{e\phi(V_b^{\ast\ast})}{|E|}\right)\left(\frac{\hbar v_D k_x}{E}\right)^2,
\end{equation}
accounting for the fact that $v_y-v_y^\prime$ is nonzero due to $e\phi(V_b^{\ast\ast})$ and increases from zero symmetrically about $k_x=0$.  Almost nesting crossing points of the dispersions of tip and sample exist only for $\phi<\phi(V_b^{\ast\ast})$ (and thus $V_b<V_b^{\ast\ast}$).
At energy $E$, these crossing points satisfy 
\begin{equation}
    \frac{\phi - \phi(V_b^{\ast\ast})}{\phi(V_b^{\ast\ast})} \sim \left(\frac{\hbar v_D k_x}{E}\right)^2,
\end{equation}
so that 
\begin{equation}
    k_\mathrm{max} \sim \frac{eV_b}{\hbar v_D} \left(\frac{|\phi - \phi(V_b^{\ast\ast})|}{\phi(V_b^{\ast\ast})}\right)^{1/2}.
\end{equation}
These relations combined with Eq.\ (\ref{eq:currint}) reproduce the parametric dependences in Eq.\ (\ref{eq:finvastast}).

The scattering approach is readily extended to inelastic current flow between tip and sample. Here, we consider the contribution of the interlayer electron-phonon coupling. The contribution of the intralayer electron-phonon coupling can be obtained by replacing the relevant electron-phonon coupling strength along the lines sketched in Sec.\ \ref{sec:qual_inelastic} above. 

The plane-wave factor $\frac{1}{\sqrt{\Omega}}e^{i\mathbf{Q}\cdot\mathbf{r}}$ of the phonon introduces the phonon wavevector $\mathbf{Q}$ into the momentum-conservation factors. Moreover, we account for the phonon energy $\hbar\omega_\mathbf{Q}$ in the energy balance and use that in addition to the interlayer tunneling amplitude $w$, the characteristic strength of the interlayer electron-phonon coupling is controlled by the ratio $\ell_\mathrm{ZPM}/a$ of the atomic zero-point motion $\ell_\mathrm{ZPM}$ and the graphene bond length $a$ (see Sec.\ \ref{sec:phonon} for a detailed discussion). We can then estimate
\begin{eqnarray}
    && T_{k_x,k_x^\prime}(E,E') \sim \frac{1}{N}\sum_{\mathbf{Q}} \frac{w^2}{\hbar^2v_y v_y^\prime}\left(\frac{\ell_\mathrm{ZPM}}{a}\right)^2 
    \nonumber\\
    && \quad \times \delta_{k_x,k_x^\prime+Q_x} L_y\delta(k_y-k_y^\prime-Q_y)\delta(E-E'-\hbar\omega_\mathbf{Q}),\qquad
\end{eqnarray}
with $N=\Omega/\Omega_\mathrm{uc}$ being the number of unit cells of area $\Omega_\mathrm{uc}$. 

For sufficiently large twist angle, $q_0\gg \kappa_F$, we approximate the phonon frequency $\omega_\mathbf{Q}\simeq \omega_{\mathbf{q}_0}$ [Fig.\ \ref{fig:smallthetaBZ}(d)]. We can then perform the sum over $\mathbf{Q}$ to obtain
\begin{equation}
    T_{k_x,k_x^\prime}(E,E') \sim \frac{L_y^2}{N} \frac{w^2}{\hbar^2v_y v_y^\prime}\left(\frac{\ell_\mathrm{ZPM}}{a}\right)^2 \delta(E-E'-\hbar\omega_{\mathbf{q}_0}).
\end{equation}
For bias voltages close to $eV_b=\hbar\omega_{\mathbf{q}_0}$, we have
\begin{eqnarray}
  && \frac{d}{dV_b} f_\mu(E)[1-f_{-\mu}(E')] \delta(E-E'-\hbar\omega_{\mathbf{q}_0}) 
    \nonumber\\
   && \qquad \qquad
     \simeq e\theta(eV_b -\hbar\omega_{\mathbf{q}_0})\delta(E-\mu)\delta(E'+\mu).\quad
\end{eqnarray}
The $\delta$-functions constrain the energies of the initial and final states to the Fermi circles of tip and sample. Thus, we can approximate $|v_y v_y^\prime| \sim v_D^2$ and the sums over $k_x$ and $k_x^\prime$ each contribute a factor of the order of the number of channels, $2\kappa_F L_x$. Collecting factors into Eq.\ (\ref{eq:currentstart}), this reproduces Eqs.\ (\ref{eq:phonon1stT}) and (\ref{eq:phonon1stL}) up to the twist-angle dependence, which we dropped in the estimate of the electron-phonon coupling. 

\section{Twisted graphene layers: Electronic states}
\label{sec:electronic}

For completeness and for fixing notation, we briefly review  some elements of the electronic properties of graphene \cite{CastroNeto2009} and of tunneling between twisted layers \cite{Bistritzer2011}. We also include a discussion of the electrostatics of graphene-graphene junctions in the QTM \cite{Inbar2023}. 

\subsection{Graphene}
\label{sec:elgraph}

\begin{figure}[b!]
    \centering
    \includegraphics[width=\columnwidth]{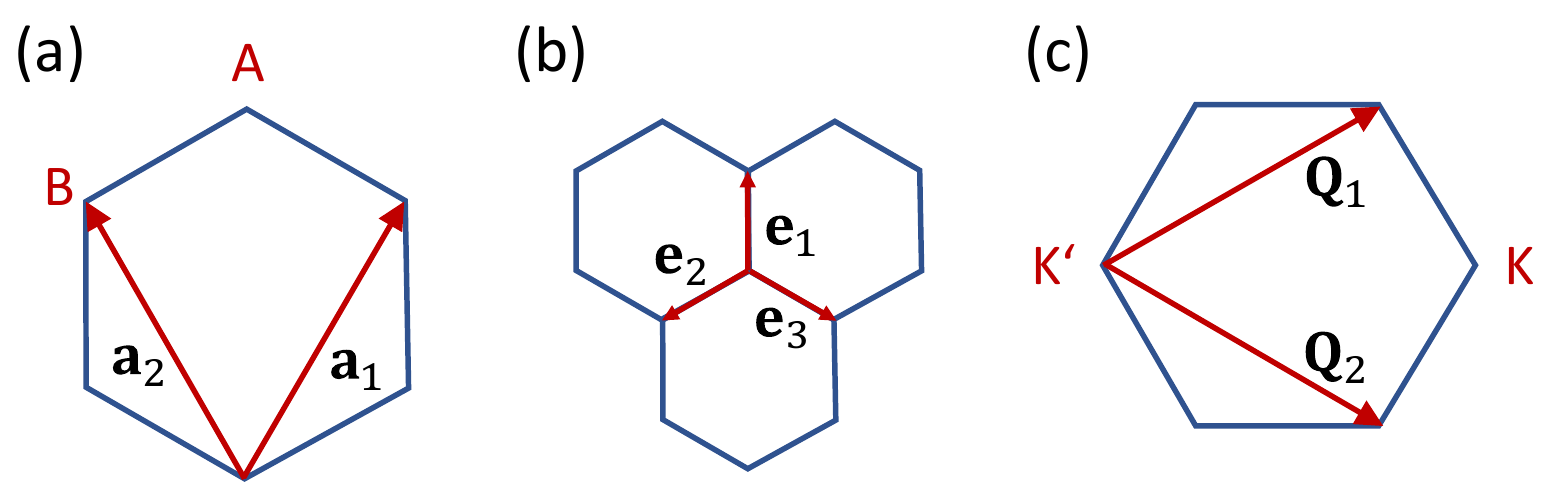}
    \caption{Honeycomb lattice of graphene layers. (a) A and B sublattice and lattice vectors $\mathbf{a}_{1/2}$ of direct lattice. (b) Bond vectors $\mathbf{e}_j$ of direct lattice.
    (c) Brillouin zone with $\mathbf{K}$ and $\mathbf{K}'$ points and reciprocal lattice vectors $\mathbf{Q}_{1/2}$. We also occasionally use $\mathbf{Q}_0=\mathbf{0}$.}
    \label{fig:lattice}
\end{figure}

Each of the two graphene layers is described by a tight-binding Hamiltonian  
\begin{equation}
    H = -t^\parallel \sum_\mathbf{R} \sum_{j=1}^3 \left\{ \ket{\mathbf{R}}\bra{\mathbf{R}+\mathbf{e}_j}
    + \ket{\mathbf{R}+\mathbf{e}_j}\bra{\mathbf{R}}
    \right\},
\end{equation}
where the $\mathbf{e}_j$ denote the three bond vectors  
\begin{equation}
    \mathbf{e}_1 = a\left(\begin{array}{c}
         0  \\
         1
    \end{array}
    \right) \quad ; \quad 
    \mathbf{e}_{2/3}=a \left(\begin{array}{c}
         \mp \sqrt{3}/2  \\
         -1/2
    \end{array}
    \right)
\end{equation}
emanating from an A site to the three nearest-neighbor B sites (Fig.\ \ref{fig:lattice}). The sum  is over the sites $\mathbf{R}$ of sublattice A. In the Bloch basis defined through
\begin{eqnarray}
    \ket{\mathbf{k},\alpha} = \frac{1}{\sqrt{N}} \sum_\mathbf{R}  e^{i\mathbf{k} \cdot(\mathbf{R}+ \boldsymbol{\tau}_\alpha)} \ket{\mathbf{R}+ \boldsymbol{\tau}_\alpha} 
    \label{eq:planewave}
\end{eqnarray}
($\alpha=A,B$ denotes the sublattice, $\boldsymbol{\tau}_A=0$, and $\boldsymbol{\tau}_B=\mathbf{e}_1$), the Hamiltonian takes the form
\begin{equation}
    H_\mathbf{k} = \left( \begin{array}{cc}
      0   &  -t^\parallel \sum_j e^{i\mathbf{k}\cdot \mathbf{e}_j} \\
      -t^\parallel \sum_j e^{-i\mathbf{k}\cdot \mathbf{e}_j}   & 0
    \end{array} \right).
\end{equation} 
The band structure
\begin{equation}
    E_{\mathbf{k},\pm} = 
    \pm t^\parallel \sqrt{A_\mathbf{k}^2 + B_\mathbf{k}^2}
\end{equation}
has valence ($-$) and conduction ($+$) bands with the lattice-periodic part 
$\ket{\mathbf{k},\pm}=(u_{A,\mathbf{k},\pm},u_{B,\mathbf{k},\pm})^T$ of the Bloch functions given by
\begin{equation}
   u_{A,\mathbf{k},\pm} = \frac{1}{\sqrt{2}} \quad ; \quad u_{B,\mathbf{k},\pm} = \mp \frac{1}{\sqrt{2}} e^{-i\gamma_\mathbf{k}}. 
   \label{eq:bwfgra}
\end{equation}
Here, we define $A_\mathbf{k}=\sum_j \cos(\mathbf{k}\cdot \mathbf{e}_j)$ and $B_\mathbf{k}=\sum_j \sin(\mathbf{k}\cdot \mathbf{e}_j)$ as well as the phase $\gamma_\mathbf{k}=\arctan (B_\mathbf{k}/A_\mathbf{k})$. For graphene 
lattice vectors $\mathbf{a}_{1/2}=a[ 
\pm \sqrt{3}/2 ,
3/2]$,  the reciprocal lattice is spanned by the vectors (see Fig.\ \ref{fig:lattice})
\begin{equation}
    \mathbf{Q}_{1/2}=\frac{4\pi}{3a} \left(\begin{array}{c}
          \sqrt{3}/2  \\
         \pm 1/2
    \end{array}
    \right).
\end{equation}
We occasionally find it useful to define $\mathbf{Q}_0=\mathbf{0}$. 

We measure wavevectors $\mathbf{k}$ from the $\Gamma$-point. For states close to the $K$-point at $\mathbf{K}=\frac{4\pi}{3a}(1/\sqrt{3},0)$ or the $K'$-point at $-\mathbf{K}$, 
the electron dispersion simplifies to the Dirac form $E_{{\boldsymbol{\kappa}},\pm} = \pm \hbar v_D|\boldsymbol{\kappa}|$ (with $v_D=3t^\parallel a/2\hbar$), where $\boldsymbol{\kappa}$ is measured from the $K$ or $K'$ point, respectively. The phase $\gamma_\mathbf{k}$ becomes 
\begin{equation}
\gamma_{\mathbf{k}} = \pi - \arctan(\kappa_y/\kappa_x)
\label{eq:phasedirac}
\end{equation}
($K$-point) and  $\gamma_{\mathbf{k}} =  \arctan(\kappa_y/\kappa_x)$ ($K'$-point).

\subsection{Tunneling between twisted layers}
\label{sec:tunneling}

We describe tunnneling between the twisted graphene layers on tip (layer 1, unprimed) and sample (layer 2, primed) following Bistrizer and MacDonald \cite{Bistritzer2011}. Starting with the Bernal-stacked configuration [see Fig.\ \ref{fig:smallthetaBZ}(a)], the sites of the A sublattice of the twisted layers (denoted $\mathbf{R}$ and $\mathbf{R}'$, respectively) are related by
\begin{equation}
    \mathbf{R}' = D(\theta) (\mathbf{R}-\mathbf{e}_1)+\mathbf{d}.
\end{equation}
Here, $D(\theta)$ is a rotation matrix involving the twist angle $\theta$ and $\mathbf{d}$ a relative shift of the rotated lattices. Note that with our conventions, the vector $\mathbf{K}'$ denotes the location of the $K$-point of the primed (sample) layer. 

Tunneling between the layers is described by the matrix elements  
\begin{equation}
    \langle \mathbf{R}+\boldsymbol{\tau}_\alpha | H_T | \mathbf{R}'+\boldsymbol{\tau}_\beta^\prime \rangle=t^\perp(\mathbf{R}+\boldsymbol{\tau}_\alpha - \mathbf{R}'-\boldsymbol{\tau}_\beta^\prime),
\end{equation}
of the (first-quantized) tunneling Hamiltonian $H_T$, where $t^\perp(\mathbf{r})$ is assumed to be only a function of the distance of the sites projected into the graphene plane.

We consider the tunneling matrix elements
\begin{equation}
    T^{\alpha\beta}_{\mathbf{k}\mathbf{p}'} = 
    \langle \mathbf{k}\alpha|H_T|\mathbf{p}'\beta \rangle
\end{equation}
between states $\ket{\mathbf{k}\alpha}$ 
with momentum $\mathbf{k}$ and sublattice $\alpha$ in the tip [Eq.\ (\ref{eq:planewave})] and states
\begin{equation}
    \ket{\mathbf{p}'\beta} = 
    \frac{1}{\sqrt{N}}
    \sum_{\mathbf{R}'}
    e^{i\mathbf{p}'\cdot (\mathbf{R}'+\boldsymbol{\tau}_\beta^\prime)} \ket{\mathbf{R}'+\boldsymbol{\tau}_\beta^\prime}
\end{equation}
with momentum $\mathbf{p}'$ and sublattice $\beta$ in the sample. The vector $\boldsymbol{\tau}_\beta^\prime$ is rotated relative to $\boldsymbol{\tau}_\beta$ by the twist angle $\theta$. Inserting definitions and expanding
\begin{equation}
t^\perp(\mathbf{r})=\frac{1}{\Omega}\sum_\mathbf{q}t^\perp_\mathbf{q}e^{i\mathbf{q}\cdot\mathbf{r}}\end{equation} 
into a Fourier series (note that the sum over $\mathbf{q}$ is not restricted to the Brillouin zone), one finds 
\begin{align}
        T^{\alpha\beta}_{\mathbf{k}\mathbf{p}'} &= 
    \frac{1}{\Omega}\sum_\mathbf{q}\frac{1}{N}
    \sum_{\mathbf{R}}
    \sum_{\mathbf{R}'}
     t^\perp_\mathbf{q}     \nonumber\\
     & \times
    e^{i\mathbf{q}\cdot(\mathbf{R}+\boldsymbol{\tau}_\alpha - \mathbf{R}'-\boldsymbol{\tau}_\beta^\prime)}
    e^{-i\mathbf{k}\cdot (\mathbf{R}+\boldsymbol{\tau}_\alpha)} e^{i\mathbf{p}'\cdot (\mathbf{R}'+\boldsymbol{\tau}_\beta^\prime)}.
\end{align}
We transform the sums over $\mathbf{R}$ and $\mathbf{R}'$ into sums over reciprocal lattice vectors $\mathbf{G}$ and $\mathbf{G}'$ of the two layers using  
\begin{equation}
\sum_\mathbf{R} e^{i\mathbf{q}\cdot\mathbf{R}}= N \sum_\mathbf{G} \delta_{\mathbf{q},\mathbf{G}} 
\end{equation}
and 
\begin{equation}
\sum_{\mathbf{R}'} e^{-i\mathbf{q}\cdot\mathbf{R}'}= N 
\sum_{\mathbf{G}} e^{-i\mathbf{G}'\cdot(-\mathbf{e}_1^\prime + \mathbf{d})}\delta_{\mathbf{q},\mathbf{G}'}.
\end{equation}
Unlike the sum over $\mathbf{R}$, the sum over $\mathbf{R}'$ generally does not include a term with $\mathbf{R}=0$, resulting in the phase factor on the right hand side. This yields \cite{Bistritzer2011}
\begin{align}
    T^{\alpha\beta}_{\mathbf{k}\mathbf{p}'} &= \sum_{\mathbf{G}_1}
    \sum_{\mathbf{G}_2}
     \frac{t^\perp_{\mathbf{k}+\mathbf{G}_1}}{\Omega_\mathrm{uc}}
     \nonumber\\
    &\times 
    e^{i \mathbf{G}_1\cdot \boldsymbol{\tau}_\alpha-i\mathbf{G}_2\cdot (\boldsymbol{\tau}_\beta
    -\mathbf{e}_1)
    -i\mathbf{G}_2^\prime\cdot \mathbf{d}}
    \delta_{\mathbf{k}+\mathbf{G}_1,\mathbf{p}'+\mathbf{G}_2^\prime}.  
    \label{eq:tunabkp}
\end{align}    
With momenta measured relative to the $\Gamma$-point, translation invariance enforces that tunneling conserve crystal momentum  modulo reciprocal lattice vectors of the two layers. 

Unlike for elastic tunneling, inelastic tunneling at larger twist angles involves virtual states far from the Fermi energy. However, matrix elements of $H_T$ always involve one momentum close to the $K$-points of one of the layers. This can be used to simplify the tunneling matrix element since $t_\mathbf{q}$ decays rapidly on the scale of the Brillouin zone \cite{Bistritzer2011}. 
Assuming for definiteness that $\mathbf{k}$ is close to the $K$-point, one retains only those terms in the sum over $\mathbf{G}_1$, in  which the momenta in $t^\perp_{\mathbf{k}+\mathbf{G}_1}$
have the smallest magnitude, i.e., the three contributions with $\mathbf{G}_1 =\mathbf{0}$,
$\mathbf{G}_1 = -\mathbf{Q}_1$, and $\mathbf{G}_1 = -\mathbf{Q}_2$ corresponding to vectors $\mathbf{k}-\mathbf{Q}_j$ located near the three equivalent $K$-points in the hexagonal Brillouin zone. Using that momentum conservation effectively restricts $\mathbf{G}_1 = \mathbf{G}_2$ for relevant twist angles, this gives
\begin{equation}
T^{\alpha\beta}_{\mathbf{k}\mathbf{p}'} \simeq w \sum_{j=0}^2 e^{-i
\mathbf{Q}_j\cdot \boldsymbol{\tau}_\alpha + i\mathbf{Q}_j\cdot (\boldsymbol{\tau}_\beta
    -\mathbf{e}_1)
    + i \mathbf{Q}_j^\prime\cdot \mathbf{d}}
    \delta_{\mathbf{k}-\mathbf{Q}_j,\mathbf{p}'-\mathbf{Q}_j^\prime}  
\end{equation}
where $w=t_\mathbf{K}^\perp/\Omega_\mathrm{uc}$. Explicit evaluation of the exponentials gives
\begin{equation}
T^{\alpha\beta}_{\mathbf{k}\mathbf{p}'} \simeq \sum_{j=0}^2 T_j^{\alpha\beta} e^{i \mathbf{Q}_j^\prime\cdot \mathbf{d}} \delta_{\mathbf{k}-\mathbf{Q}_j,\mathbf{p}'-\mathbf{Q}_j^\prime}  
    \label{eq:mael}
\end{equation}
with the matrices
\begin{equation}
    T_0 = w  
    \left(\begin{array}{cc}
      1 & 1 \\ 1 & 1  
    \end{array}\right) \quad , \quad T_{1/2} = w \left(\begin{array}{cc}
      e^{\mp i\zeta} & 1 \\ e^{\pm i\zeta} & e^{\mp i\zeta} 
    \end{array}\right) 
    \label{eq:T}
\end{equation}
in sublattice space. Here, we used the abbreviation $\zeta = 2\pi/3$. We note that the same expression holds when $\mathbf{p}'$ is close to the $K$-point of the sample, but $\mathbf{k}$ is possibly further from the tip's $K$-point.

\subsection{Matrix elements}
\label{sec:tunneling2}

Calculations of the tunneling current by Fermi's golden rule require the matrix elements of the tunneling Hamiltonian between eigenstates of the upper and lower layers,
\begin{equation}  T^{ss'}_{\mathbf{k}\mathbf{p}'} =\langle \mathbf{k},s|H_T|\mathbf{p}',s'\rangle.
\end{equation}
Here, $s,s'=\pm$ enumerate the valence and conduction bands of the top and bottom graphene layers, respectively. According to Eq.\ (\ref{eq:mael}), we can write 
\begin{equation}
T^{ss'}_{\mathbf{k}\mathbf{p}'} \simeq \sum_{j=0}^2 T^{ss'}_{\mathbf{k}\mathbf{p}';j} e^{i \mathbf{Q}_j^\prime\cdot \mathbf{d}} \delta_{\mathbf{k}-\mathbf{Q}_j,\mathbf{p}'-\mathbf{Q}_j^\prime}.  
\label{eq:sumjtj}
\end{equation}
with 
\begin{equation}  T^{ss'}_{\mathbf{k}\mathbf{p}';j} =\langle \mathbf{k},s|T_j|\mathbf{p}',s'\rangle.
\label{eq:Tsskpj}
\end{equation}
Equation (\ref{eq:T}) gives 
\begin{equation}
    \langle u|{T}_j|u'\rangle
    = w e^{-ij\zeta}(u_A + e^{ij\zeta}u_B)^* (u_A^\prime + e^{ij\zeta} u_B^\prime),
\end{equation}
where we used $e^{3i\zeta}=1$. Interestingly the matrix elements factorize into independent contributions of the two spinors.  Using the explicit Bloch spinors in Eq.\ (\ref{eq:bwfgra}), this becomes
\begin{equation}
    T^{ss'}_{\mathbf{k}\mathbf{p}';j} = 
  \frac{w}{2} e^{-ij\zeta} \left[1 - s e^{-i(\gamma_\mathbf{k}-j\zeta)}\right]^* \left[1 - s' e^{-i(\gamma^\prime_{\mathbf{p}'}-j\zeta)}\right].\label{eq:matrixDirac0}
\end{equation}
We note that the phases 
$\gamma_{\mathbf{k}}$ and 
$\gamma^\prime_{\mathbf{p}'}$ are defined in terms of the bond vectors of tip and sample, respectively. We can also express the matrix elements for momenta 
$\mathbf{k}$ and $\mathbf{p}'$ close to the Dirac points. Writing $\mathbf{k}=\mathbf{K}+\boldsymbol{\kappa}$ and $\mathbf{p}'=\mathbf{K}'+\boldsymbol{\pi}'$, specifying to the $K$-point, and using Eq.\ (\ref{eq:phasedirac}), we find 
\begin{equation}
    T^{ss'}_{\boldsymbol{\kappa}\boldsymbol{\pi}';j} = 
  \frac{we^{-ij\zeta}}{2}  \!\left[1 + s e^{i(\theta_{\boldsymbol{\kappa}}-\frac{\theta}{2}+j\zeta)}\right] \!\left[1 + s' e^{i(\theta_{\boldsymbol{\pi}'}+\frac{\theta}{2}+j\zeta)}\right].
  \label{eq:matrixDirac}
\end{equation}
Note that here,  we have defined the angles $\theta_{\boldsymbol{\kappa}}$ and $\theta_{\boldsymbol{\pi}'}$ in a global coordinate system, relative to which the tip/sample layers are rotated by $\pm \theta/2$.   

\subsection{Electrostatics}
\label{sec:elst}

We review the electrostatics of the QTM contact \cite{Inbar2023}. We assume a configuration (see Fig.\ \ref{fig:elst}), in which a bias voltage $V_b$ is applied between tip and sample. The electron densities $n_\mathrm{T}$ and $n_\mathrm{S}$ of tip and sample are further controlled by gate voltages  $V_\mathrm{TG}=V_G+V_D$ and $V_\mathrm{BG}=V_G-V_D$ applied to the top and bottom gates. Here, we defined the symmetrized and antisymmetrized gate voltages $V_\mathrm{G} = \frac{1}{2}(V_\mathrm{TG} + V_\mathrm{BG})$ and $V_\mathrm{D}= V_\mathrm{TG}-V_\mathrm{BG}$. 

The gate electrodes are assumed to have a high density of states (large quantum capacitance), so that their chemical potentials are independent of the applied voltages. We set $\mu_\mathrm{TG}\simeq \mu_\mathrm{BG}\simeq 0$. Consequently, the gate voltages control their electric potentials, 
\begin{equation}
    e(V_\mathrm{G}+V_\mathrm{D}) = e\phi_\mathrm{TG} \quad ; \quad e(V_\mathrm{G}-V_\mathrm{D}) = e\phi_\mathrm{BG}.  
\end{equation}
Similarly, the bias voltage $V_b$ controls the electrochemical potentials (encompassing the chemical potentials $\mu$ and the electrostatic potentials $\phi$) of tip (T) and sample (S), 
\begin{eqnarray}
    \pm\frac{eV_b}{2} = \mu_\mathrm{T/S} +e\phi_\mathrm{T/S} \label{eq:Vbias1}
\end{eqnarray}
The chemical potentials are related to the electron densities of tip and sample through
\begin{equation}
   n_\mathrm{T/S} = N_f \frac{\mu^2_\mathrm{T/S}}{4\pi \hbar^2 v_D^2}\,\mathrm{sgn}\mu_\mathrm{T/S}.
   \label{eq:nmu}
\end{equation}
Here, $N_f=4$ is the number of flavors and we assume that the graphene dispersions can be approximated as linear for relevant densities. 

Electrostatics relates the potentials and electron densities through 
\begin{eqnarray}
    e(\phi_\mathrm{TG}-\phi_\mathrm{T}) &=& -\frac{e^2 d_g}{\epsilon_\mathrm{hBN}\epsilon_0}n_\mathrm{TG} \\
    e(\phi_\mathrm{T}-\phi_\mathrm{S}) &=& \frac{e^2 d}{\epsilon\epsilon_0}(n_\mathrm{S} + n_\mathrm{BG}) 
    \nonumber\\
    &=& -\frac{e^2 d}{\epsilon\epsilon_0}(n_\mathrm{T} + n_\mathrm{TG})\\
    e(\phi_\mathrm{S}-\phi_\mathrm{BG}) &=& \frac{e^2 d_g}{\epsilon_\mathrm{hBN}\epsilon_0}n_\mathrm{BG}.
\end{eqnarray}
Here, $d_g$ denotes the thickness of the gate dielectrics (dielectric constant $\epsilon_\mathrm{hBN}$) and $d$ the distance of tip and sample (dielectric constant $\epsilon$). Finally, the relation
\begin{equation}
   n_\mathrm{T} + n_\mathrm{S} =  -(n_\mathrm{TG} + n_\mathrm{BG}).
   \label{eq:densities}
\end{equation}
is imposed by overall charge neutrality. 

We first solve these equations for the charges on tip and sample. The gate voltage $V_\mathrm{G}$ controls the overall charge density in tip and sample, 
\begin{equation}
    2eV_\mathrm{G} = \frac{e^2 d_g}{\epsilon_\mathrm{hBN}\epsilon_0}(n_\mathrm{T} + n_\mathrm{S}) - (\mu_\mathrm{T} + \mu_\mathrm{S}).
\end{equation}
Assuming that the screening lengths are small compared to $d_g$ in gate electrodes as well as tip and sample, we can further neglect the chemical potential shifts, so that 
\begin{equation}
    2eV_\mathrm{G} = \frac{e^2 d_g}{\epsilon_\mathrm{hBN}\epsilon_0}(n_\mathrm{T} + n_\mathrm{S}).
\end{equation}
Correspondingly, the difference in electron densities on tip and sample is controlled by the bias voltage $V_b$ in conjunction with the displacement field $V_\mathrm{D}$,
\begin{equation}   \frac{\epsilon_\mathrm{hBN}\epsilon_0}{e^2 d_g}2eV_\mathrm{D} - \frac{2\epsilon\epsilon_0}{e^2 d} eV_b 
= \frac{2\epsilon\epsilon_0}{e^2 d}(\mu_\mathrm{T}-\mu_\mathrm{S})  + (n_\mathrm{T}-n_\mathrm{S}).
\end{equation}
Here, we assumed that $d\ll d_g$. Equation (\ref{eq:nmu}) can now be used to extract $n_\mathrm{T/S}$ as well as $\mu_\mathrm{T/S}$. This in turn yields $\phi_\mathrm{T/S}$ with Eq.\ (\ref{eq:Vbias1}) . We note that the setup in Fig.\ \ref{fig:elst} admits independent control of the chemical potentials $\mu_\mathrm{T}$ and $\mu_\mathrm{S}$ as well as the potential difference $\phi_\mathrm{T}-\phi_\mathrm{S}$.

In our analytical calculations, we choose $V_\mathrm{G}=V_\mathrm{D}=0$ and a small quantum capacitance. Then, tip and sample are overall charge neutral,
$n_\mathrm{T}+n_\mathrm{S}=0$, and  have opposite chemical potentials, $\mu_\mathrm{T} = -\mu_\mathrm{S} = \mu$. For a linear graphene dispersion, the density of states per flavor at the (bias-dependent) Fermi energy is $\nu(\mu) = \mu/(2\pi\hbar^2 v_D^2)$ for both layers and the difference $ \phi = \phi_\mathrm{T} - \phi_\mathrm{S}$ in electric potentials is
\begin{equation}
    e\phi = \frac{e^2d}{\epsilon\epsilon_0}n_\mathrm{T} = (q_\mathrm{TF}d) \mu \ll \mu,
    \label{eq:phimuTF}
\end{equation}
where $q_\mathrm{TF} = (e^2/2\epsilon\epsilon_0) N_f\nu(\mu)$ is the Thomas-Fermi wavevector. Thus, $eV_b = 2\mu + e\phi \simeq 2\mu$. At the same time, $e\phi$ gives a relative shift of the Dirac points in tip and sample. This leads to the bias regimes sketched in Fig.\ \ref{fig:smalltheta}.

The electrostatics of the contact region ($d\ll d_g$) differs from that far from the contact ($d\gg d_g$), where the gate charges directly control the electron densities of tip and sample, $n_\mathrm{T}\simeq -n_\mathrm{TG}$ and $n_\mathrm{S}\simeq -n_\mathrm{BG}$. For certain parameters, this difference in electrostatics induces $pn$-junctions in the sample, which enclose the contact region. This leads to Fabry-Perot-type resonances within the sample, which affect the measured tunneling currents. We do not consider this experimental issue in the following, as it can be avoided in phonon spectroscopy by a judicious choice of parameters.

\section{Elastic tunneling}
\label{sec:eltun}

We begin our discussion of the tunneling conductance of QTM junctions by considering elastic tunneling. For small twist angles, the Dirac cones of tip and sample layers intersect at small bias voltages and for the same valley, as illustrated in Fig.\ \ref{fig:smallthetaBZ}(c). 
In practice, one limits the strong tunnel coupling in this limit by separating tip and sample by a few atomic layers of a transition metal dichalcogenide (e.g., WSe$_2$) \cite{Inbar2023}.

Complementing the scattering approach sketched in Sec.\ \ref{sec:overview}, we evaluate the current from tip to sample using Fermi's golden rule,
\begin{align}
    I & =  \frac{2\pi e N_f}{\hbar}
    \sum_{s,s'}
    \sum_\mathbf{k}\sum_{\mathbf{p}'} 
    |\langle \mathbf{k},s|H_T|\mathbf{p}',s'\rangle|^2
    \nonumber\\ &\times \delta(E_{\mathbf{k},s}^{(\mathrm{T})} + e\phi_\mathrm{T} - E_{\mathbf{p}',s'}^{(\mathrm{S})}-e\phi_\mathrm{S} )
    \nonumber\\
    &\times \left[f_{\mu_\mathrm{T}}(E_{\mathbf{k},s}^{(\mathrm{T})}) -f_{\mu_\mathrm{S}}(E_{\mathbf{p}',s'}^{(\mathrm{S})})\right].
\end{align}
For overall charge neutrality and small quantum capacitance (see Sec.\ \ref{sec:elst}), the tunneling current becomes
\begin{eqnarray}
    I &=&  \frac{2\pi e N_f}{\hbar}
    \sum_{s,s'}
    \sum_\mathbf{k}\sum_{\mathbf{p}'} 
    |\langle \mathbf{k},s|H_T|\mathbf{p}',s'\rangle|^2
    \nonumber\\
    &&\times\delta(E_{\mathbf{k},s}^{(\mathrm{T})} + e\phi - E_{\mathbf{p}',s'}^{(\mathrm{S})} )
    \nonumber\\
    &&\times\left[f_{eV_b/2}(E_{\mathbf{k},s}^{(\mathrm{T})}) -f_{-eV_b/2}(E_{\mathbf{p}',s'}^{(\mathrm{S})})\right].
\end{eqnarray}
In describing elastic scattering at small twist angles, it is advantageous to measure momenta from the $K$-points of tip and sample. With $\mathbf{k}=\mathbf{K}+\boldsymbol{\kappa}$ and $\mathbf{p}'=\mathbf{K}'+\boldsymbol{\pi}'$ as well as Eq.\ (\ref{eq:sumjtj}) for the tunneling matrix elements, we find 
\begin{eqnarray}
    I &=&  \frac{2\pi e N_f}{\hbar}
    \sum_{s,s'}
    \sum_{\boldsymbol{\kappa}}\sum_{\boldsymbol{\pi}'} 
    \sum_j |T_{\boldsymbol{\kappa},\boldsymbol{\pi}';j}^{ss'}|^2
    \delta_{\boldsymbol{\pi}'-\boldsymbol{\kappa},\mathbf{q}_j}
    \nonumber\\
    &&\times 
    \delta(E_{\boldsymbol{\kappa},s} + e\phi - E_{\boldsymbol{\pi}',s'})
    \nonumber\\
    && \times\left[f_{eV_b/2}(E_{\boldsymbol{\kappa},s}) -f_{-eV_b/2}(E_{\boldsymbol{\pi}',s'})\right].
    \label{eq:Ielastic}
\end{eqnarray}
We dropped the layer superscripts on the dispersions, which are identical provided that trigonal warping can be neglected. 

One expects structure in the differential conductance $dI/dV_b$ at two characteristic voltages  \cite{Inbar2023}, see Fig.\ \ref{fig:smalltheta}. As the bias voltage increases, the tunneling current onsets at 
\begin{equation}
eV_b^*=\hbar v_D q_j = 
2\hbar v_D|\mathbf{K}|\sin\frac{\theta}{2},
\label{eq:vstar}
\end{equation} 
which depends linearly on small twist angles. At a larger bias voltage [see Eq.\ (\ref{eq:phimuTF}) and Fig,\ \ref{fig:smalltheta}] 
\begin{equation}
    V_b^{**} = \frac{2 V_b^*}{q_\mathrm{TF}d} \gg V_b^*, 
\end{equation}
the potential difference $\phi$ leads to nesting of the Dirac cones of tip and sample. This occurs when
\begin{equation}
e\phi (V_b^{**})=\hbar v_D q_j = 
2\hbar v_D|\mathbf{K}|\sin\frac{\theta}{2}.
\end{equation}
The linear dependence of $V_b^*$ on (small) $\theta$ implies that $V_b^{**}$ has a leading square-root dependence. Retaining the electric potential in 
the relation $eV_b=2\mu + e\phi$ yields a subleading linear term.

\subsection{Voltages of order \texorpdfstring{$V_b^*$}{of the first characteristic voltage}}
\label{sec:Vast}

We first consider voltages of order $V_b^*$, such that the electrostatic potential difference $\phi$ is negligible.  The energy $\delta$-function imposes $s=s'$, so that 
\begin{eqnarray}
    I &=&  \frac{2\pi e N_f }{\hbar}
    \sum_j\sum_{s}
    \sum_{\boldsymbol{\kappa}} 
    |T_{\boldsymbol{\kappa},\boldsymbol{\kappa}+\mathbf{q}_j;j}^{ss}|^2
\delta(E_{\boldsymbol{\kappa},s}  - E_{\boldsymbol{\kappa}+\mathbf{q}_j,s})
    \nonumber\\
    &&\times
    \left[f_{eV_b/2}(E_{\boldsymbol{\kappa},s}) -f_{-eV_b/2}(E_{\boldsymbol{\kappa}+\mathbf{q}_j,s})\right].
\end{eqnarray}
It is convenient to rewrite this as \begin{eqnarray}
    I &=&  \frac{2\pi e N_f }{\hbar}    \sum_j\sum_{\boldsymbol{\kappa}} \sum_s
    |T_{\boldsymbol{\kappa},\boldsymbol{\kappa}+\mathbf{q}_j;j}^{ss}|^2
    \nonumber\\
    &&\times
    |E_{\boldsymbol{\kappa},s}  + E_{\boldsymbol{\kappa}+\mathbf{q}_j,s}|
\delta(E^2_{\boldsymbol{\kappa},s}  - E^2_{\boldsymbol{\kappa}+\mathbf{q}_j,s})
    \nonumber\\
    &&\times
    \left[f_{eV_b/2}(E_{\boldsymbol{\kappa},s}) -f_{-eV_b/2}(E_{\boldsymbol{\kappa}+\mathbf{q}_j,s})\right].
\end{eqnarray}
We pass to the differential conductance at zero temperature. The derivative of the first Fermi function places $E_{\boldsymbol{\kappa},s}$ and thus also $E_{\boldsymbol{\kappa}+\mathbf{q}_j,s}$ at the chemical potential $\mu = eV_b/2$ with $s=+$. Corresponding results hold for the derivative of the second Fermi function,   
which yields  
\begin{eqnarray}
    && \frac{d I}{d V_b} =  \frac{2\pi e^3 N_f V_b}{\hbar}    \sum_j\sum_{\boldsymbol{\kappa}}\sum_s 
    |T_{\boldsymbol{\kappa},\boldsymbol{\kappa}+\mathbf{q}_j;j}^{ss}|^2  \nonumber\\
    && \qquad \times
\delta(E^2_{\boldsymbol{\kappa},+}  - E^2_{\boldsymbol{\kappa}+\mathbf{q}_j,+}) \delta(E_{\boldsymbol{\kappa},+}-\frac{eV_b}{2})
 .
\end{eqnarray}
The first $\delta$-function enforcing energy conservation has the argument 
\begin{equation}
E^2_{\boldsymbol{\kappa},+}  - E^2_{\boldsymbol{\kappa}+\mathbf{q}_j,+}
= \hbar^2v_D^2 q_j (2\kappa \cos\theta_j + q_j),
\end{equation}
where $\theta_j = \measuredangle (\boldsymbol{\kappa},\mathbf{q}_j)$. 
Thus, the $\delta$-function imposes $\theta_j\simeq \pi$ for bias voltages near the threshold, which implies $\theta_{\boldsymbol{\kappa}} + j\zeta = -\pi/2$ and $\theta_{\boldsymbol{\kappa}+\mathbf{q}_j} + j\zeta = \pi/2$. Evaluating the matrix element in Eq.\ (\ref{eq:matrixDirac}) for this case gives 
$ |T_{\boldsymbol{\kappa},\boldsymbol{\kappa}+\mathbf{q}_j;j}^{ss}|^2 = w^2$
for small twist angles, which is independent of $j$ and $s$. Thus, we find 
\begin{eqnarray}
    && \frac{d I}{d V_b} \simeq  \frac{12\pi e^3 N_f w^2 V_b}{\hbar}  
    \, \Omega  \nu({eV_b}/{2}) 
    \nonumber\\
    &&  \times
     \frac{1}{\hbar^2 v_D^2 q_0} \int \frac{d\theta_0}{2\pi}
\delta(2\kappa_F \cos\theta_0 + q_0). \qquad
\end{eqnarray}
Performing the angular integral yields the result given in Eq.\ (\ref{eq:finvast}) of Sec.\ \ref{sec:overview}.

\subsection{Voltages of order \texorpdfstring{$V_b^{**}$}{of the second characteristic voltage}}
\label{sec:Vastast}

In the vicinity of $V_b^{**}$, we need to retain the electrostatic potential in Eq.\ (\ref{eq:Ielastic}). We write the potential as $\phi = \phi(V_b^{**}) + \delta\phi$ and evaluate the current for small $\delta\phi$. For small quantum capacitance,  $q_\mathrm{TF}d\ll 1$, the chemical potential $\mu=eV_b/2$ is much larger than $\phi$, so that we retain only the contributions of regions I and III in Fig.\ \ref{fig:smalltheta}(c), where tunneling satisfies $s=s'$. Regions I and III give identical contributions by particle-hole symmetry. Moreover, the sum over $j$ gives a factor of three in view of $C_3$ symmetry. Thus, we have
\begin{eqnarray}
    I &=&  \frac{12\pi e N_f}{\hbar}
    \sum_{\boldsymbol{\kappa}}|T_{\boldsymbol{\kappa},\boldsymbol{\kappa}+\mathbf{q}_0;0}^{++}|^2\delta(E_{\boldsymbol{\kappa},+} + e\phi - E_{\boldsymbol{\kappa}+\mathbf{q}_0,+})
    \nonumber\\
    && \times\left[f_{eV_b/2}(E_{\boldsymbol{\kappa},+}) -f_{-eV_b/2}(E_{\boldsymbol{\kappa}+\mathbf{q}_0,+})\right].
\end{eqnarray}
The Fermi-function factor is nonzero and equal to unity as long as $0<\hbar v_D\kappa<eV_b/2$. 

For small $\delta\phi$, we can approximate
\begin{equation}
   E_{\boldsymbol{\kappa},+} + e\phi - E_{\boldsymbol{\kappa}+\mathbf{q}_0,+} 
   \simeq e\delta\phi + \hbar v_D \frac{\kappa q_0(1-\cos\theta_0)}{\kappa + q_0},
   \label{eq:argdelta}
\end{equation}
We thus have $\theta_0 \ll 1$. Moreover, we observe that $\theta_{\boldsymbol{\kappa}}\simeq\theta_{\boldsymbol{\kappa}+\mathbf{q}_0}\simeq \pi/2$ in region I. According to Eq.\ (\ref{eq:matrixDirac}), we can now approximate the matrix element as $|T_{\boldsymbol{\kappa},\boldsymbol{\kappa}-\mathbf{q}_0;0}^{++}|^2\simeq w^2$ for small twist angles. Thus, we find 
\begin{eqnarray}
    I &=&  \frac{3 e N_f w^2 \Omega}{\pi\hbar}\int_{0}^{\mu/(\hbar v_D)}
    d\kappa \kappa 
    \nonumber\\
    &&\times \int d\theta_0 \,
\delta\!\left(e\delta\phi +  \frac{\hbar v_D\kappa q_0}{2(\kappa +  q_0)}\theta_0^2\right)
    .\label{eq:Iastast}
\end{eqnarray}
We note that the integral over $\kappa$ is dominated by the upper limit, so that we can approximate $\kappa+q_0\simeq \kappa$ in the argument of the $\delta$-function. Evaluating the remaining integrals yields the result in Eq.\ (\ref{eq:finvastast}) of Sec.\ \ref{sec:overview}. 

\section{Phonons and electron-phonon coupling}
\label{sec:phonon}

\subsection{Mode expansion}

We consider the phonon modes of tip or sample layer, neglecting mechanical coupling between the graphene layers. 
We expand the atomic displacements $\mathbf{u}(\mathbf{R}+\boldsymbol{\tau}_\alpha)$ of each of the layers, including both in-plane and out-of-plane components, 
into phonon modes (annihilation operator $b_{r,\mathbf{Q}}$) enumerated by their momenta $\mathbf{Q}$ and mode index $r$,  
\begin{align}
   &\mathbf{u}(\mathbf{R}+\boldsymbol{\tau}_\alpha,t) = \frac{1}{\sqrt{N}} \sum_\mathbf{Q}\sum_r
   \boldsymbol{\epsilon}_{r,\mathbf{Q}}^\alpha
   \frac{1}{\sqrt{2M \omega_{r,\mathbf{Q}}}} 
   \nonumber\\
   &\qquad\times e^{i\mathbf{Q}\cdot (\mathbf{R}+\boldsymbol{\tau}_\alpha)} \left(b_{r,\mathbf{Q}} e^{-i\omega_{r,\mathbf{Q}}t} + b^\dagger_{r,-\mathbf{Q}} e^{i\omega_{r,\mathbf{Q}}t}
   \right).
   \label{eq:modeexp}
\end{align}    
Here, $M$ is the mass of the unit cell and the momentum sum is restricted to the Brillouin zone. The polarization vectors $\boldsymbol{\epsilon}_{r,\mathbf{Q}}^\alpha$ denote the mode displacement of sublattice $\alpha$, which we normalize according to 
\begin{equation}
    \sum_\alpha M_\alpha (\boldsymbol{\epsilon}^\alpha_{r,\mathbf{q}})^\ast \cdot \boldsymbol{\epsilon}^\alpha_{r',\mathbf{q}} = M \delta_{r,r'}
\end{equation}   with the $M_\alpha$ denoting the mass of the atom on sublattice $\alpha$ (i.e., $M_\alpha = M/2$ for graphene layers). We keep the mode expansion of the atomic displacements general throughout this paper to facilitate generalization from the case of graphene to other types of layers on tip and sample. 

For graphene, the polarization vectors can be chosen real due to inversion symmetry. The polarization vectors satisfy the general relation 
\begin{equation}
[\boldsymbol{\epsilon}_{r,\mathbf{Q}}^\alpha]^* = \boldsymbol{\epsilon}_{r,-\mathbf{Q}}^\alpha,
\end{equation}
while inversion symmetry implies $\boldsymbol{\epsilon}_{r,\mathbf{Q}}^\alpha = \boldsymbol{\epsilon}_{r,-\mathbf{Q}}^\alpha$. 
At long wavelengths, the phonon modes include longitudinal and transverse acoustic modes, longitudinal and transverse optical modes, as well as a flexural mode. The quadratic dispersion of the flexural mode is cut off at a nonzero frequency at long wavelengths due to the coupling between the graphene layers and the substrates, even in the absence of mechanical coupling between tip and sample. 

\subsection{Electron-phonon coupling}

Electron-phonon coupling arises from changes in the bond lengths of the graphene layers associated with the atomic displacements as well as from the deformation potential. Changes in the intralayer bond lengths result in the electron-phonon coupling of individual graphene layers. At long wavelengths, this intralayer coupling can be incorporated in the Dirac description as a gauge field. The atomic displacements also modify the interlayer tunneling discussed in Sec.\ \ref{sec:tunneling}, leading to interlayer electron-phonon coupling. The electron-phonon coupling originating from the deformation potential gives a separate contribution to the intralayer coupling. 

\subsubsection{Intralayer electron-phonon coupling}

We consider the intralayer electron-phonon due to changes in the bond lengths 
for one of the layers. The atomic displacements modify the electronic tight-binding Hamiltonian of the graphene layer by  
\begin{equation}
    H_\mathrm{intra} = - \sum_{\mathbf{R}} \sum_{i=1}^3 \delta t^\parallel_{\mathbf{R},\mathbf{R}+\mathbf{e}_i} \ket{\mathbf{R}}\bra{\mathbf{R}+{\mathbf{ e}}_i} + \mathrm{h.c.}
    \label{Hintra1}
\end{equation}
Assuming that the hopping amplitude $t^\parallel$ depends only on the interatomic distance, one finds 
\begin{equation}
   \delta t^\parallel_{\mathbf{R},\mathbf{R}+\mathbf{e}_i} \simeq \beta  \mathbf{\hat e}_{i}\cdot [\mathbf{u}(\mathbf{R} + \mathbf{e}_i)-\mathbf{u}(\mathbf{R})]
   \label{Hintra2}
\end{equation}
to linear order in the atomic displacements $\mathbf{u}$. Here, $\mathbf{\hat e}_{j}$ denotes the unit vector in the direction of $\mathbf{e}_{j}$ and 
$\beta={\partial t^\parallel}/{\partial a}$ the derivative of the hopping amplitude with respect to the bond length. Notice that the intralayer coupling is limited to in-plane phonons. Coupling to flexural phonons appears only in quadratic order in the atomic displacements and can be neglected, since there is coupling to flexural modes already at linear order in the interlayer electron-phonon coupling. 

The matrix elements of the electron-phonon coupling can be obtained by inserting the mode expansion in Eq.\ (\ref{eq:modeexp}) into $H_\mathrm{intra}\ket{\mathbf{k},A}$. This yields
\begin{widetext}
\begin{equation}
    H_\mathrm{intra}\ket{\mathbf{k},A}
    = -\frac{1}{N} \sum_\mathbf{R} \sum_j   \sum_\mathbf{Q}\sum_r
     \frac{{\beta}}{\sqrt{2M \omega_{r,\mathbf{Q}}}}  \mathbf{\hat e}_{j}\cdot 
[\boldsymbol{\epsilon}_{r,\mathbf{Q}}^B e^{i\mathbf{Q}\cdot \mathbf{e}_j} - \boldsymbol{\epsilon}_{r,\mathbf{Q}}^A]\left(b_{r,\mathbf{Q}} e^{-i\omega_{r,\mathbf{Q}}t} + b^\dagger_{r,-\mathbf{Q}} e^{i\omega_{r,\mathbf{Q}}t}
   \right) e^{i(\mathbf{k}+\mathbf{Q})\cdot\mathbf{R}}
    \ket{\mathbf{R}+\mathbf{e}_j}.
\end{equation}
Evaluating the sum over $\mathbf{R}$ yields

\begin{equation}    
H_\mathrm{intra}\ket{\mathbf{k},A}
    = -\frac{1}{\sqrt{N}} \sum_j   \sum_\mathbf{Q}\sum_r
     \frac{{\beta}}{\sqrt{2M \omega_{r,\mathbf{Q}}}} \mathbf{\hat e}_{j}\cdot  [\boldsymbol{\epsilon}_{r,\mathbf{Q}}^B
 - \boldsymbol{\epsilon}_{r,\mathbf{Q}}^A
e^{-i\mathbf{Q}\cdot \mathbf{e}_j}]\left(b_{r,\mathbf{Q}} e^{-i\omega_{r,\mathbf{Q}}t} + b^\dagger_{r,-\mathbf{Q}} e^{i\omega_{r,\mathbf{Q}}t}
   \right) e^{-i\mathbf{k}\cdot \mathbf{e}_j}
    \ket{\mathbf{k}+\mathbf{Q},B}
\end{equation}
with $\ket{\mathbf{k}+\mathbf{Q},B}$ interpreted as the state, for which $\mathbf{k}+\mathbf{Q}$ is folded back into the Brillouin zone. Here, we have used the identity
$ \frac{1}{\sqrt{N}} \sum_\mathbf{R} e^{i\mathbf{p}\cdot \mathbf{R}} \ket{\mathbf{R}+\mathbf{e}_j} = e^{-i \mathbf{p}\cdot \mathbf{e}_j} \ket{\mathbf{p},B}$, which can be checked by direct calculation. Similarly, one finds
\begin{eqnarray}
   H_\mathrm{intra}\ket{\mathbf{k},B}
    =  - \frac{1}{\sqrt{N}} \sum_j   \sum_\mathbf{Q}\sum_r
    \frac{{\beta}}{\sqrt{2M \omega_{r,\mathbf{Q}}}} \mathbf{\hat e}_{j}\cdot  [
 \boldsymbol{\epsilon}_{r,\mathbf{Q}}^B e^{i\mathbf{Q}\cdot \mathbf{e}_j} -\boldsymbol{\epsilon}_{r,\mathbf{Q}}^A
]\left(b_{r,\mathbf{Q}} e^{-i\omega_{r,\mathbf{Q}}t} + b^\dagger_{r,-\mathbf{Q}} e^{i\omega_{r,\mathbf{Q}}t}
   \right) e^{i\mathbf{k}\cdot \mathbf{e}_j} | \mathbf{k}+\mathbf{Q},A \rangle.
\end{eqnarray} 
As a result, the intralayer electron-phonon interaction takes the form 
\begin{equation}
    H_\mathrm{intra} = \sum_\mathbf{Q} \sum_r
    \left(b_{r,-\mathbf{Q}} e^{-i\omega_{r,-\mathbf{Q}}t} + b^\dagger_{r,\mathbf{Q}} e^{i\omega_{r,\mathbf{Q}}t}
   \right) \sum_\mathbf{k} \left\{
    \ket{\mathbf{k}-\mathbf{Q},B}
    M_{\mathbf{k}-\mathbf{Q},B;\mathbf{k};A}^r  \bra{\mathbf{k},A} + \ket{\mathbf{k}-\mathbf{Q},A} [M_{\mathbf{k},B;\mathbf{k}-\mathbf{Q};A}^r]^* \bra{\mathbf{k},B} \right\}
   \label{eq:helphintra}
\end{equation}
\end{widetext}
with the electron-phonon matrix element
\begin{align}
    & M_{\mathbf{k}-\mathbf{Q},B;\mathbf{k},A}^r  = \frac{1}{\sqrt{N}} \sum_j  
     \frac{{\beta}}{\sqrt{2M \omega_{r,\mathbf{Q}}}}
     \nonumber\\ & \qquad  \times \mathbf{\hat e}_{j}\cdot  \left[
\boldsymbol{\epsilon}_{r,\mathbf{Q}}^A
e^{-i(\mathbf{k}-\mathbf{Q})\cdot \mathbf{e}_j} - \boldsymbol{\epsilon}_{r,\mathbf{Q}}^B e^{-i\mathbf{k}\cdot \mathbf{e}_j} \right].
\end{align}
For acoustic phonons, the term in square brackets vanishes linearly in $\mathbf{Q}$, while the denominator vanishes only as $|\mathbf{Q}|^{1/2}$. Thus, the intralayer electron-phonon coupling vanishes in the long-wavelength limit. For optical phonons, the coupling approaches a constant in the long-wavelength limit. 

\subsubsection{Interlayer electron-phonon coupling }

The interlayer electron-phonon coupling arises from modifications in the interlayer distances between atoms in the two layers. Importantly, this contribution to the electron-phonon coupling cannot be obtained starting with the continuum model of twisted bilayer graphene, even at long phonon wavelengths. The reason is that as we will see, the dominant contribution to the coupling arises from phonon-induced changes of phases of the tunneling matrix element, which depend on the large momenta of the K-points as measured from the $\Gamma$-point.

The interlayer tunneling amplitude is a function of the in-plane and out-of-plane distances of the atoms in the two layers. Including the modification of the distances by the atomic displacements, we have 
\begin{eqnarray}
   &&t^\perp(|\mathbf{R}+\boldsymbol{\tau}_\alpha - \mathbf{R}'-\boldsymbol{\tau}_\beta^\prime+\mathbf{u}_\parallel-\mathbf{u}_\parallel^\prime|,d+u_\perp-u_\perp^\prime)   \quad \\
  && \quad = \frac{1}{\Omega}\sum_{\mathbf{q}} t^\perp_\mathbf{q}(d+u_\perp-u_\perp^\prime) e^{i\mathbf{q}\cdot(\mathbf{R}+\boldsymbol{\tau}_\alpha - \mathbf{R}'-\boldsymbol{\tau}_\beta^\prime+\mathbf{u}_\parallel-\mathbf{u}_\parallel^\prime)}.\nonumber
\end{eqnarray}
Here, we used the shorthands $\mathbf{u} = \mathbf{u}(\mathbf{R}+\boldsymbol{\tau}_\alpha)$ and $\mathbf{u}^\prime = \mathbf{u}^\prime(\mathbf{R}^\prime +\boldsymbol{\tau}_\alpha^\prime)$ and denote the equilibrium interlayer distance by $d$. Expanding to linear order in the displacements yields
\begin{widetext}
\begin{equation}
 t^\perp(|\mathbf{R}+\boldsymbol{\tau}_\alpha - \mathbf{R}'-\boldsymbol{\tau}_\beta^\prime+\mathbf{u}_\parallel-\mathbf{u}_\parallel^\prime|,d+u_\perp-u_\perp^\prime)      \simeq\frac{1}{\Omega}\sum_{\mathbf{q}} t^\perp_\mathbf{q}(d) e^{i\mathbf{q}\cdot (\mathbf{R}+\boldsymbol{\tau}_\alpha - \mathbf{R}'-\boldsymbol{\tau}_\beta^\prime)} \{ 1 + i\mathbf{q}\cdot (\mathbf{u}_\parallel-\mathbf{u}_\parallel^\prime) 
 + \frac{\partial \ln t^\perp_\mathbf{q}(d)}{\partial d}(u_\perp-u_\perp^\prime) \} .
\end{equation} 
The first term in the curly brackets is just the interlayer tunneling discussed in Sec.\ \ref{sec:tunneling} \cite{Bistritzer2011}. The contributions of the second and third terms, collectively denoted $\delta t^\perp$, encapsulate the electron-phonon coupling  
\begin{equation}
    H_\mathrm{inter} = \sum_\mathbf{R}\sum_{\mathbf{R}'} \sum_{\alpha,\beta}\ket{\mathbf{R}+ \boldsymbol{\tau}_\alpha} \delta t^\perp(|\mathbf{R} + \boldsymbol{\tau}_\alpha -\mathbf{R}'-\boldsymbol{\tau}_\beta^\prime +\mathbf{u}_\parallel-\mathbf{u}_\parallel^\prime|,d+u_\perp-u_\perp^\prime)
    \bra{\mathbf{R}'+\boldsymbol{\tau}_\beta^\prime} + \mathrm{h.c.}
\end{equation}
\end{widetext}
to in-plane ($\mathbf{u}_\parallel$) and out-of-plane ($u_\perp$) phonons.

The derivation of matrix elements $\langle \mathbf{k},\alpha | H_\mathrm{inter} | \mathbf{p}',\beta \rangle$ closely follows the discussion of interlayer tunneling in Sec.\ \ref{sec:tunneling}. Using the mode expansion in Eq.\ (\ref{eq:modeexp}) of the atomic displacements, one finds the following rule for shifting the electronic momenta in the expressions for the tunneling matrix element:  Scattering from phonons in the tip (wavevector $\mathbf{Q}$) can be accounted for by shifting the outgoing momentum according as $\mathbf{k}\to \mathbf{k}-\mathbf{Q}$ and leaves the incoming momentum $\mathbf{p}'$ invariant. Scattering from phonons in the sample (wavevector $\mathbf{Q}'$) can be accounted for by shifting the incoming momentum $\mathbf{p}'\to \mathbf{p}'+\mathbf{Q}'$ and leaves the outgoing momentum $\mathbf{k}$ invariant. This yields
\begin{widetext}
\begin{equation}
       \langle \mathbf{k},\alpha | H_\mathrm{inter} | \mathbf{p}',\beta \rangle
    = \sum_\mathbf{Q} \sum_r 
    \langle  \mathbf{k},\alpha|H_\mathrm{inter}|\mathbf{p}',\beta;r,\mathbf{Q}\rangle \left(b_{r,\mathbf{Q}} e^{-i\omega_{r,\mathbf{Q}}t} + b^\dagger_{r,-\mathbf{Q}} e^{i\omega_{r,\mathbf{Q}}t}   \right) - (\mathbf{Q}\to \mathbf{Q}')  ,
\end{equation}
where [cp.\ Eq.\ (\ref{eq:tunabkp})]
\begin{eqnarray}
    \langle  \mathbf{k},\alpha|H_\mathrm{inter}|\mathbf{p}',\beta;r,\mathbf{Q}\rangle
    &=&  \frac{1}{\sqrt{N}}\sum_{\mathbf{G}_1}
    \sum_{\mathbf{G}_2}
    e^{i \mathbf{G}_1\cdot \boldsymbol{\tau}_\alpha-i\mathbf{G}_2\cdot (\boldsymbol{\tau}_\beta
    -\mathbf{e}_1)
    -i\mathbf{G}_2^\prime\cdot \mathbf{d}} \, \delta_{\mathbf{k}-\mathbf{Q} + \mathbf{G}_1,\mathbf{p}'+\mathbf{G}_2^\prime}
   \nonumber\\
    && \times \frac{1}{\sqrt{2M \omega_{r,\mathbf{Q}}}}\left\{ \frac{t^\perp_{\mathbf{p}'+\mathbf{G}_2^\prime}}{\Omega_\mathrm{uc}}  i(\mathbf{p}'+\mathbf{G}_2^\prime)\cdot \boldsymbol{\epsilon}^\alpha_{r,\mathbf{Q}}
 + \frac{1}{\Omega_\mathrm{uc}}\frac{\partial  t^\perp_{\mathbf{p}'+\mathbf{G}_2^\prime}(d)}{ \partial d}  \mathbf{\hat z}\cdot \boldsymbol{\epsilon}^\alpha_{r,\mathbf{Q}}
    \right\}  
    \label{eq:ephmatel1}
\end{eqnarray} 
and 
\begin{eqnarray}
    \langle  \mathbf{k},\alpha|H_\mathrm{inter}|\mathbf{p}',\beta;r,\mathbf{Q}'\rangle
    &=& - \frac{1}{\sqrt{N}}\sum_{\mathbf{G}_1}
    \sum_{\mathbf{G}_2}
    e^{i \mathbf{G}_1\cdot \boldsymbol{\tau}_\alpha-i\mathbf{G}_2\cdot (\boldsymbol{\tau}_\beta
    -\mathbf{e}_1)
    -i\mathbf{G}_2^\prime\cdot \mathbf{d}} \, \delta_{\mathbf{k} + \mathbf{G}_1,\mathbf{p}'+\mathbf{Q}' + \mathbf{G}_2^\prime} 
   \nonumber\\
    &&  \times \frac{1}{\sqrt{2M \omega_{r,\mathbf{Q}'}}}\left\{  \frac{t^\perp_{\mathbf{k}+\mathbf{G}_1}}{\Omega_\mathrm{uc}}  i(\mathbf{k}+\mathbf{G}_1)\cdot \boldsymbol{\epsilon}^\beta_{r,\mathbf{Q}'}
+ \frac{1}{\Omega_\mathrm{uc}}\frac{\partial t^\perp_{\mathbf{k}+\mathbf{G}_1}(d)}{ \partial d}\mathbf{\hat z}\cdot \boldsymbol{\epsilon}^\beta_{r,\mathbf{Q}'}
    \right\}  .
    \label{eq:ephmatel2}
\end{eqnarray} 
\end{widetext}
The first terms in the curly brackets in Eqs.\ (\ref{eq:ephmatel1}) and (\ref{eq:ephmatel2}) describe coupling to in-plane phonons, while the second terms describe coupling to flexural phonons. 

The contribution of $H_\mathrm{inter}$ to the inelastic tunneling current can be accounted for in first-order perturbation theory. Consequently, the electronic momenta $\mathbf{k}$ and $\mathbf{p}^\prime$ are close to the $K$-points of tip and sample. In this case, we can further simplify the matrix elements. As discussed in Sec.\ \ref{sec:tunneling}, the dominant terms are then given by $\mathbf{G}_1=0$, $\mathbf{G}_1 = -\mathbf{Q}_1$, and $\mathbf{G}_1=-\mathbf{Q}_2$ with $\mathbf{G}_1=\mathbf{G}_2$. Moreover, except for the smallest twist angles, we can neglect the distance of $\mathbf{k}$ and $\mathbf{p}^\prime$ from the respective $K$-points relative to the distance $q_j$ between the $K$-points of tip and sample. With these approximations, the Kronecker-$\delta$  imposing momentum conservation 
in Eq.\ (\ref{eq:ephmatel1}) simplifies as $\delta_{\mathbf{k}-\mathbf{Q} + \mathbf{G}_1,\mathbf{p}'+\mathbf{G}_2^\prime} \to \delta_{\mathbf{K}-\mathbf{Q} - \mathbf{Q}_j,\mathbf{K}'-\mathbf{Q}_j^\prime} = \delta_{\mathbf{Q,\mathbf{q}_j}}$, so that the phonon wavevector equals a connection vector between the Dirac points of tip and sample 
[and analogously for Eq.\ (\ref{eq:ephmatel2})]. Thus, we find  
\begin{widetext}
\begin{eqnarray}
    \langle  \mathbf{k},\alpha|H_\mathrm{inter}|\mathbf{p}',\beta;r,\mathbf{Q}\rangle
    &=&  \frac{1}{\sqrt{N}}\sum_{j=0}^2 T_j^{\alpha\beta} e^{
    i\mathbf{Q}_j^\prime\cdot \mathbf{d}} \, \delta_{\mathbf{Q},\mathbf{q}_j}
    \frac{1}{\sqrt{2M \omega_{r,\mathbf{Q}}}}\left\{ i(\mathbf{K}'-\mathbf{Q}_j^\prime)\cdot \boldsymbol{\epsilon}^\alpha_{r,\mathbf{Q}}
 + \frac{\tilde w}{w}   \mathbf{\hat z}\cdot \boldsymbol{\epsilon}^\alpha_{r,\mathbf{Q}}
    \right\},  
    \label{eq:ephmatel11} \\
    \langle  \mathbf{k},\alpha|H_\mathrm{inter}|\mathbf{p}',\beta;r,\mathbf{Q}'\rangle
    &=& - \frac{1}{\sqrt{N}}\sum_{j=0}^2 T_j^{\alpha\beta}
    e^{i\mathbf{Q}_j^\prime\cdot \mathbf{d}} \, \delta_{\mathbf{Q}', \mathbf{q}_j} 
 \frac{1}{\sqrt{2M \omega_{r,\mathbf{Q}'}}}\left\{ i(\mathbf{K}-\mathbf{Q}_j)\cdot \boldsymbol{\epsilon}^\beta_{r,\mathbf{Q}'}
+ \frac{\tilde w}{w}\mathbf{\hat z}\cdot \boldsymbol{\epsilon}^\beta_{r,\mathbf{Q}'}
    \right\}  .
    \label{eq:ephmatel22}
\end{eqnarray} 
\end{widetext}
Here, we defined $\tilde w=\frac{\partial w}{\partial d}$. (Note that $\tilde w$ and $w$ have different dimensions.) Several comments are in order concerning these results in the limit of small twist angles. (i) The vectors $\mathbf{q}_j$  and hence the phonon wavevector $\mathbf{Q}$ are approximately perpendicular to $\mathbf{K}-\mathbf{Q}_j$ as well as $\mathbf{K}^\prime -\mathbf{Q}_j^\prime$. Thus, as a consequence of the scalar product, the coupling to in-plane phonons is predominantly to transverse phonon modes for the phonon vectors probed in QTM experiments. (ii) The coupling to transverse acoustic modes diverges as $|\mathbf{Q}|^{-1/2}$ at small phonon wavevectors $\mathbf{Q}$. (iii) The coupling to the transverse acoustic modes is effectively to layer-antisymmetric phonons, also known as the phason mode.

As mentioned in the beginning of this section, the relevant momenta entering Eqs.\ (\ref{eq:ephmatel11}) and (\ref{eq:ephmatel22}) are the large momenta of the K-points rather than the phonon wavevector. It is for this reason that even in the long-wavelength limit, this coupling cannot be obtained by adding phonon displacements to the continuum model of twisted bilayer graphene. Instead, one has to follow the derivation of the continuum model after taking the phonon displacements into account.

The underlying reason is that the interlayer coupling arises from phase factors associated with the interlayer tunneling. We use this observation in
App.\ \ref{app:interlayer} to show that by means of a gauge transformation, the interlayer coupling can be brought into a form, which is analogous to the intralayer coupling. As it appears in this section, the contribution of the interlayer coupling to the inelastic tunneling current can be accounted for in a first-order golden rule calculation. In the transformed form, it must be treated in second order. Remarkably, in the transformed form, the intra- and interlayer couplings are related by the replacement
\begin{equation}
    \frac{\partial t^\parallel}{\partial a} \leftrightarrow i t_\parallel \mathbf{K}.
\end{equation}
This shows that one expects both contributions to the inelastic tunneling current to have corresponding parametric dependences, as born out by the explicit calculations, see Table \ref{tab:elph}.

\subsubsection{Deformation-potential coupling}
\label{sec:deformation}

For long-wavelength phonons, the deformation potential 
\begin{equation}
    V(\mathbf{r}) = -D \boldsymbol{\nabla}\cdot \mathbf{u}
    \label{def_potential}
\end{equation}
gives an additional contribution to the electron-phonon coupling of longitudinal acoustic phonons. This can be compared with the intralayer electron-phonon coupling in Eqs.\ (\ref{Hintra1}) and (\ref{Hintra2}) due to changes in the bond length. In contrast to the deformation potential, the latter is of comparable magnitude for transverse and longitudinal acoustic phonons. 
Apart from this difference, the magnitudes of the deformation-potential and the gauge coupling are related by the replacement
\begin{equation}
    D \leftrightarrow \frac{\partial t^\parallel}{\partial \ln a}.
\end{equation}
We can use this correspondence to 
obtain the additional contribution of the deformation potential to the inelastic tunneling current from the result for the intralayer gauge coupling. 

We also note that the deformation coupling locally shifts the chemical potential, inducing changes of the charge density. These charge fluctuations will be screened by electron-electron interactions, effectively reducing the strength of the  deformation potential. We take $D$ to be the renormalized coupling. 

\section{Phonon spectroscopy} 
\label{sec:inelastic}

\subsection{Inelastic tunneling current}
\label{sec:inelastic1}

We are now in a position to compute the inelastic tunneling current to the leading orders in tip-sample tunneling and electron-phonon coupling. Inelastic electron tunneling in conjunction with phonon emission emerges from the interlayer electron-phonon coupling ${\mathcal H}_\mathrm{inter}$ [see Eqs.\ (\ref{eq:ephmatel11}) and (\ref{eq:ephmatel22})] as well as the intralayer electron-phonon coupling $\mathcal{H}_\mathrm{intra}$ in conjunction with interlayer tunneling. While the first contribution can be captured by Fermi's golden rule in lowest order, the second requires a higher-order calculation. 

Keeping the calculation general at first, we assume a nonzero matrix element 
$\langle \mathbf{p}',s';r,\mathbf{Q} | \mathcal{T}_\mathrm{inel}|\mathbf{k},s\rangle$
of the inelastic contribution to the $\mathcal{T}$-matrix  without specifying to a particular process. We assume zero temperature and $eV_b>0$. Then, tunneling is unidirectional from tip to sample and only phonon emission contributes. We also specify to charge neutrality and small quantum capacitance, so that $\phi_\mathrm{T} = \phi_\mathrm{S}$ and $\mu = eV_b/2$ at bias voltages corresponding to typical phonon frequencies. With these assumptions, the phonon contribution $\delta I$ to the tunneling current from tip to sample is given by 
\begin{widetext}
\begin{equation}
    \delta I = 
    \frac{2\pi e N_f}{\hbar} \sum_{\mathbf{Q}} \sum_r\sum_{\mathbf{k},\mathbf{p}'}
    \sum_{s,s'}|\langle 
    \mathbf{p}',s';r,\mathbf{Q}|\mathcal{T}_\mathrm{inel}|\mathbf{k},s\rangle|^2
    \delta(E_{\mathbf{p}',s'}+ \hbar\omega_{r,\mathbf{Q}}-E_{\mathbf{k},s})
    f_{\mu}(E_{\mathbf{k},s})[1-f_{-\mu}(E_{\mathbf{p}',s'})]
    + (\mathbf{Q}\to\mathbf{Q}').
\end{equation}
Passing to the differential conductance in the limit of zero temperature using $\mu=eV_b/2$ gives
\begin{eqnarray}
    && \frac{d\delta I}{dV_b} = 
    \frac{2\pi e^2 N_f}{2\hbar}
    \sum_{\mathbf{Q}}
    \sum_r\sum_{\mathbf{k},\mathbf{p}'}
    \sum_{s,s'}
    \theta(eV_b-\hbar\omega_{r,\mathbf{Q}})
     |\langle \mathbf{p}',s';r,\mathbf{Q}|\mathcal{T}_\mathrm{inel}|\mathbf{k},s\rangle|^2
    \nonumber\\
     && \qquad
      \times \left\{ \delta(E_{\mathbf{p}',s'}+\hbar\omega_{r,\mathbf{Q}}-\mu)\delta(E_{\mathbf{k},s}-\mu) + 
    \delta(-\mu+\hbar\omega_{r,\mathbf{Q}}-E_{\mathbf{k},s})\delta(E_{\mathbf{p}',s'}+\mu)
 \right\} + ({\mathbf{Q}\rightarrow \mathbf{Q}')}.
\end{eqnarray}
The singular contribution to the second derivative, ${d^2\delta I}/{dV_b^2}$,  arises from the derivative of the threshold factor $\theta(eV_b-\hbar\omega_{r,\mathbf{Q}})$. The resulting $\delta$-function enforces that the bias voltage match the phonon energy. Using this constraint, the two terms in curly brackets become equal, and we obtain
\begin{equation}
    \frac{d^2\delta I}{dV_b^2}  = 
    \frac{2\pi e^3 N_f}{\hbar}
    \sum_{\mathbf{Q}}\sum_r
    \delta(eV_b-\hbar\omega_{r,\mathbf{Q}})
     \sum_{\mathbf{k},\mathbf{p}'}
    |\langle \mathbf{p}',-;r,\mathbf{Q}|\mathcal{T}_\mathrm{inel}|\mathbf{k},+\rangle|^2\delta(E_{\mathbf{p}',-}+\mu)\delta(E_{\mathbf{k},+}-\mu)+ ({\mathbf{Q}\rightarrow \mathbf{Q}')}.
\end{equation}
Here, we also used that tunneling at the threshold is from $s=+$ to $s'=-$. Due to the $\delta$-functions, the initial and final electron states are located at the Fermi energies of tip and sample, respectively. Thus, we find
\begin{equation}
    \frac{d^2\delta I}{dV_b^2}  = 
    \frac{2\pi e^3 N_f \Omega^2}{\hbar} \nu(\mu)\nu(-\mu)\sum_r\sum_{\mathbf{Q}}
    \delta(eV_b-\hbar\omega_{r,\mathbf{Q}})
    \int \frac{d\theta_\mathbf{k}}{2\pi}\frac{d\theta_{\mathbf{p}^\prime}}{2\pi} |\langle \mathbf{p}',-;r,\mathbf{Q}|\mathcal{T}_\mathrm{inel}|\mathbf{k},+\rangle|^2 + ({\mathbf{Q}\rightarrow \mathbf{Q}')},
    \label{eq:d2IdV2gen}
\end{equation}
which depends on Fermi-circle averages of the $\mathcal{T}$-matrix element. 

\subsection{Interlayer electron-phonon coupling: First-order perturbation theory}
\label{sec:inelastic2}

We first consider the contribution $\delta I_\mathrm{inter}$ of the interlayer electron-phonon coupling, $\mathcal{T}_\mathrm{inel}\to \mathcal{H}_\mathrm{inter}$ in Eq.\ (\ref{eq:d2IdV2gen}). We can readily perform the angular averages in Eq.\ (\ref{eq:d2IdV2gen}) using Eqs.\ (\ref{eq:ephmatel11}), (\ref{eq:ephmatel22}), and (\ref{eq:matrixDirac}). By $C_3$ symmetry, the sum over $j$ implicit in the matrix elements can be accounted for by a factor of three and we obtain 
\begin{eqnarray}
    &&\frac{d^2\delta I_\mathrm{inter}}{dV_b^2}  = 
    \frac{6\pi e^3 N_f \Omega^2}{\hbar N} \sum_r
    \delta(eV_b-\hbar\omega_{r,\mathbf{q}_0})\frac{\nu(\mu)\nu(-\mu)}{2M \omega_{r,\mathbf{q}_0}} 
    \nonumber\\
    &&\qquad\qquad \times 
    \sum_\alpha \left\{\left| iw\mathbf{K}'\cdot \boldsymbol{\epsilon}^\alpha_{r,\mathbf{Q}=\mathbf{q}_0}
 + \tilde w \mathbf{\hat z}\cdot \boldsymbol{\epsilon}^\alpha_{r,\mathbf{Q}=\mathbf{q}_0}
    \right|^2
    + \left|iw\mathbf{K}\cdot \boldsymbol{\epsilon}^\alpha_{r,\mathbf{Q}'=\mathbf{q}_0}
+ \tilde w\mathbf{\hat z}\cdot \boldsymbol{\epsilon}^\alpha_{r,\mathbf{Q}'=\mathbf{q}_0}
    \right|^2\right\}.
\end{eqnarray}
Here, we also used that by symmetry, $\omega_{r,\mathbf{q}_0}$ is the same for  the tip and sample layers. Finally we note that the contributions of phonon emissions in tip and sample are identical in magnitude and use the fact that the polarization vectors are real. This yields the result
\begin{equation}
    \frac{d^2\delta I_\mathrm{inter}}{dV_b^2}  = 
     G_\mathrm{incoh}
\sum_r \delta(V_b-\hbar\omega_{r,\mathbf{q}_0}/e)
     \frac{8\pi^2(\kappa_F\ell_{r,\mathbf{q}_0})^2}{\sqrt{3}}\sum_\alpha \left\{  ({\mathbf{\hat K}}'\cdot \boldsymbol{\epsilon}^\alpha_{r,\mathbf{Q}=\mathbf{q}_0})^2
 + \frac{{\tilde w}^2}{w^2|\mathbf{K}|^2} (\mathbf{\hat z}\cdot \boldsymbol{\epsilon}^\alpha_{r,\mathbf{Q}=\mathbf{q}_0})^2
    \right\}.
    \label{eq:ResultInter}
\end{equation}
\end{widetext}
Here, we have rewritten the prefactor by introducing the length $\ell_{r,\mathbf{q}_0} = \sqrt{\hbar/M \omega_{r,\mathbf{q}_0}}$  characterizing the contribution of phonon mode $r$ with wavevector $\mathbf{q}_0$ to the amplitude of the zero-point motion of the atoms and using $|\mathbf{K}|^2\Omega_\mathrm{uc} = 8\pi^2/(3\sqrt{3})$. We note that in Sec.\ \ref{sec:qual_inelastic} as well as Table \ref{tab:elph}, we use the less specific notation $\ell_\mathrm{ZPM}$ for $\ell_{r,\mathbf{q}_0}$. 

Each phonon mode contributes a $\delta$-function peak to ${d^2\delta I}/{dV_b^2}$ at $eV_b=\hbar\omega_{r,\mathbf{q}_0}$. Several comments are in order: (i) At small twist angles, transverse phonons contribute more strongly to the inelastic tunneling  current as the vectors $\mathbf{K}$ and $\mathbf{K}'$ are nearly orthogonal to the phonon wavevector $\mathbf{q}_0$. (ii) As the twist angle decreases and the phonon wavevector $q_0\to 0$, the strength of the electron-phonon coupling diverges for acoustic phonons. Concurrently, the Fermi wavevector $\kappa_F$ decreases. On balance, we have $(\kappa_F\ell_r)^2\sim \omega_{r,\mathbf{q}_0}$, so that the strength of the phonon resonance decreases as the twist angle becomes smaller. We note, however, that this is specific to the case of overall charge neutrality. Away from charge neutrality, the Fermi wavevector remains nonzero at zero bias and one finds a singular enhancement at small twist angles. The cutoff at small twist angles is discussed at the end of Sec.\ \ref{sec:qual_inelastic}. 
(iii) For the same reason, acoustic phonons contribute more weakly than optical phonons at overall charge neutrality, but stronger away from charge neutrality provided that the phonon frequencies are small compared to the zero-bias chemical potential. 

In deriving $\frac{d^2\delta I}{dV_b^2}$, we neglected the electronic momenta  measured from the respective $K$-points, relative to $q_0$ [see Eqs.\ (\ref{eq:ephmatel11}) and (\ref{eq:ephmatel22}) and the preceding discussion]. Within this approximation, all inelastic tunneling events involve phonons with momenta $\mathbf{q}_j$ and $\frac{d^2\delta I}{dV_b^2}$ becomes a $\delta$-peak as a function of bias voltage. Retaining the small electronic momenta effectively broadens the $\delta$-peak. The phonon momenta are now equal to $\mathbf{q}_j$ only up to wavevectors of order $\kappa_F$. For acoustic modes, we can estimate the change in phonon frequency as $\Delta\omega_r\sim \hbar c\kappa_F$ with $c$ the speed of sound. Thus, we find
\begin{equation}\frac{\Delta\omega_r}{\omega_{r,\mathbf{q}_0}} \sim \frac{c}{v_D} \ll 1 
\end{equation}
for the relative broadening of the phonon peak. We note that this approximation also implies a limitation on the twist angle due to the requirement $\kappa_F \ll q_0\approx |\mathbf{K}|\theta$. For an optical mode (for which this condition is more stringent), this implies
\begin{equation}
    \theta \gg \frac{c}{v_D}.
    \label{eq:twistoptlim}
\end{equation}
Here, we estimated the frequency of the  optical phonon as $c|\mathbf{K}|$.

\subsection{Intralayer electron-phonon coupling: Second-order perturbation theory}
\label{sec:inelastic3}

The contribution of the intralayer electron-phonon interaction is obtained by $\mathcal{T}_\mathrm{inel} \to
\mathcal{H}_T\mathcal{G}_0\mathcal{H}_\mathrm{intra}
+ \mathcal{H}_\mathrm{intra}\mathcal{G}_0\mathcal{H}_T$ in Eq.\ (\ref{eq:d2IdV2gen}). This is written in second quantization (as indicated by calligraphic symbols) to keep proper track of the relative exchange phase of the two contributions. 

\begin{figure*}[t]    \centering \includegraphics[width=.99\columnwidth]{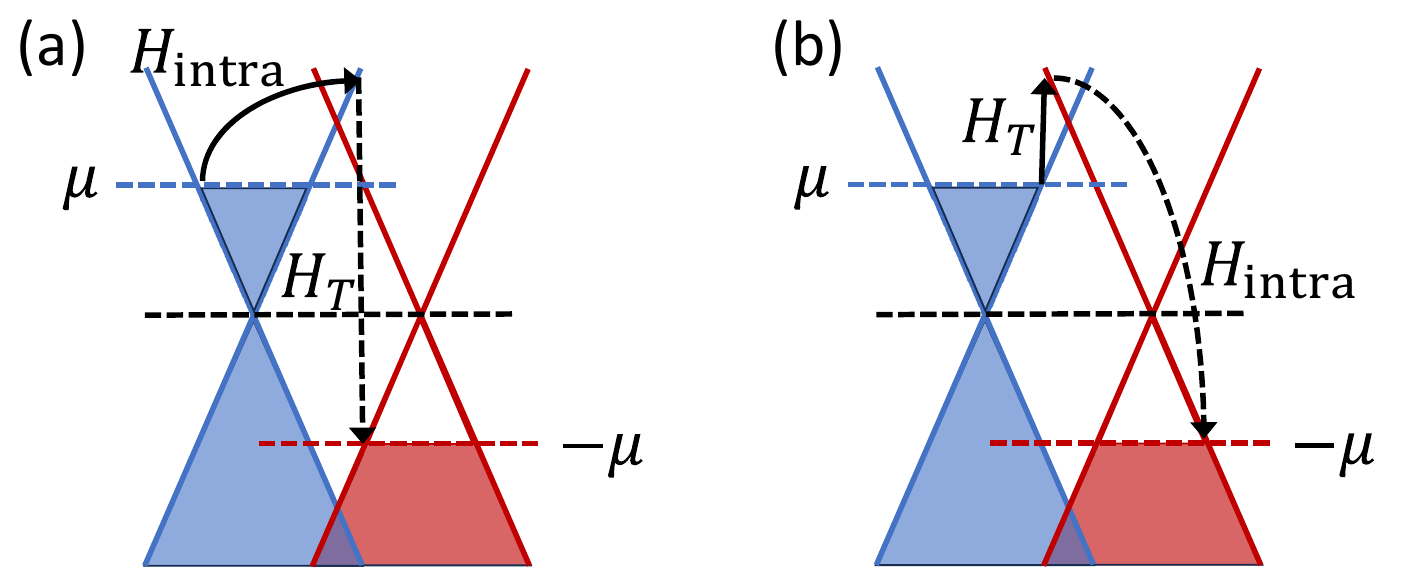} \includegraphics[width=.99 \columnwidth]{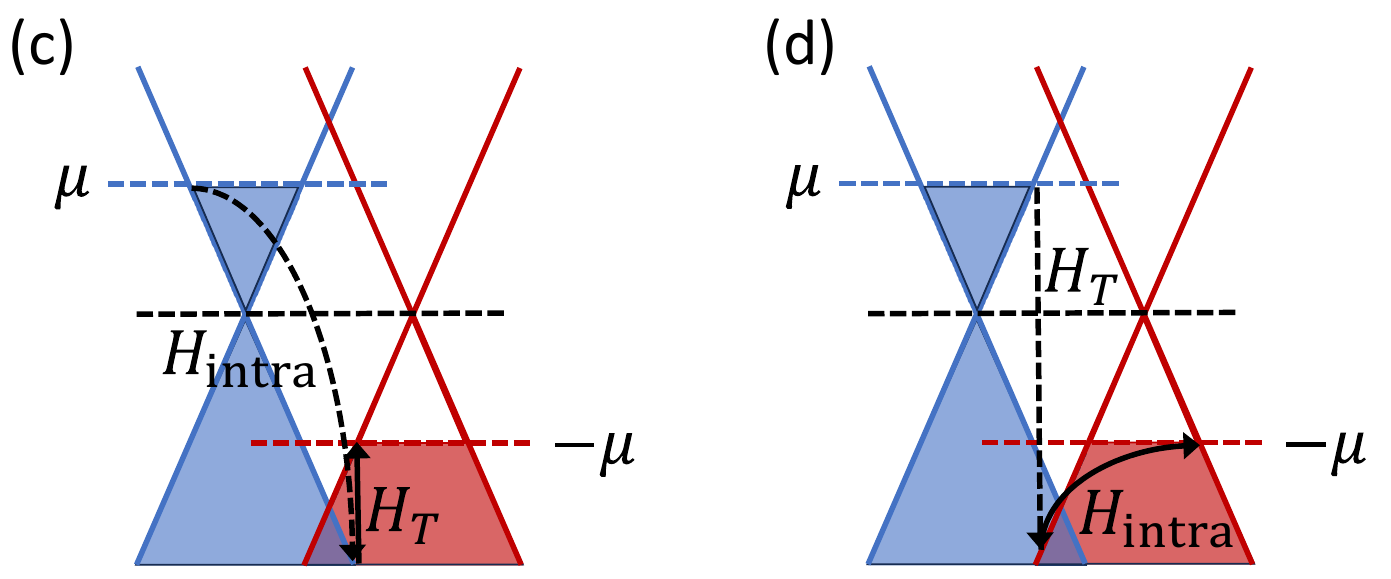} \caption{Second-order processes contributing to the inelastic tunneling current involving the intralayer electron-phonon coupling. (a,b) Processes with intermediate  electrons in the conduction band. (c,d) Processes with intermediate valence holes. All processes are shown at the threshold voltage $eV_b=\hbar\omega_{r,\mathbf{q}_j}$, with full (dashed) black lines indicating the first (second) process. Notice that due to the difference between electronic and phononic velocities, actual phonon energies (as well as $\mu$) are much smaller than represented here for reasons of readibility. This also implies that the electronic momenta measured from the respective Dirac points are much smaller than the distance between the Dirac points of tip and sample.}  \label{fig:process1}
\end{figure*}

The relevant processes are illustrated in Fig.\ 
\ref{fig:process1}, for a bias voltage at threshold ($eV_b=\hbar\omega_{r,\mathbf{q}_j}$).  Electron tunneling is from the chemical potential in the tip (conduction band) to the chemical potential in the sample (valence band). We  specify to electron wavevectors in the vicinity of $\mathbf{K}$ and $\mathbf{K}'$. Figures \ref{fig:process1}(a) and (c) describe phonon emission in the tip, (b) and (d) in the sample. By symmetry, both sets of processes contribute equally. For definiteness, we specify to emission of a phonon in the tip. 

In the process shown in Fig.\ \ref{fig:process1}(a), tunneling follows the initial phonon emission, so that the process  
contributes to $\mathcal{H}_T\mathcal{G}_0\mathcal{H}_\mathrm{intra}$. Similarly, in the process shown in Fig.\ \ref{fig:process1}(c), phonon emission follows the initial tunneling, so that the process  
contributes to $\mathcal{H}_\mathrm{intra}\mathcal{G}_0\mathcal{H}_T$. These processes are described by the matrix elements
\begin{widetext}
\begin{eqnarray}
    \langle \mathbf{p}',-;r,\mathbf{Q}| H_T G_0 H_\mathrm{intra} |\mathbf{k},+ \rangle = \sum_{\mathbf{r}}\frac{ \langle \mathbf{p}',-;r,\mathbf{Q}| H_T |\mathbf{r},+;r,\mathbf{Q}\rangle \langle \mathbf{r},+;r,\mathbf{Q}|
    H_\mathrm{intra} |\mathbf{k},+ \rangle}{E_{\mathbf{k},+}-E_{\mathbf{r},s}-\hbar\omega_{r,\mathbf{Q}}}
\label{eq:me1}
\end{eqnarray} 
and 
\begin{eqnarray}
    \langle \mathbf{p}',-;r,\mathbf{Q}| H_\mathrm{intra} G_0 H_T |\mathbf{k},+ \rangle = \sum_{\mathbf{r}}\frac{\langle \mathbf{p}',-;r,\mathbf{Q}| H_\mathrm{intra} |\mathbf{p}',-; \overline {\mathbf{r},-};\mathbf{k},+\rangle \langle \mathbf{p}',-; \overline {\mathbf{r},-};\mathbf{k},+|
    H_T |\mathbf{k},+ \rangle}{E_{\mathbf{r},s}-E^\prime_{\mathbf{p}',-}}
\label{eq:me2}
\end{eqnarray}
\end{widetext}
of the $\mathcal{T}$-matrix, respectively. Here, we denote holes by overlines. In this section, we also denote energies of tip (sample) without (with) prime. Energy and momentum  conservation demand that $E_{\mathbf{k},+}=E^\prime_{\mathbf{p}',-}+\hbar\omega_\mathbf{Q}$ as well as $\mathbf{Q}=\mathbf{k}-\mathbf{p}' \simeq \mathbf{q}_j$. 
The three contributions to tunneling are related by $C_3$ symmetry and incoherent, so that we focus on the $j=0$ contribution, which leaves the momentum unchanged, i.e., $\mathbf{r}=\mathbf{p}'$ in both processes. 

The processes in Fig.\ \ref{fig:process1}(a) and (c) contribute with a relative statistical minus sign. This follows as they 
involve 
the operator products 
$c^\dagger_{\mathbf{p},-}c_{\mathbf{r},+}c^\dagger_{\mathbf{r},+}c_{\mathbf{k},+}$ and $c^\dagger_{\mathbf{r},-}c_{\mathbf{k},+} c^\dagger_{\mathbf{p},-}c_{\mathbf{r},-}$, respectively. Anticommuting the operators in the second product turns it into $-c^\dagger_{\mathbf{p},-}c^\dagger_{\mathbf{r},-}c_{\mathbf{r},-}c_{\mathbf{k},+} $. The relative minus sign  follows by noting that $c_{\mathbf{r},+}c^\dagger_{\mathbf{r},+}$ and $c^\dagger_{\mathbf{r},-}c_{\mathbf{r},-}$ both reduce to unity when acting on the ground state. 

The  energy denominators of the two processes in Eqs.\ (\ref{eq:me1}) and (\ref{eq:me2}) are equal to each other up to corrections which are small in the ratio of the phonon and electron velocities.
As can also be seen by inspection of Figs.\ \ref{fig:process1}(a) and (c), the energy denominators differ by the phonon energy $\hbar\omega_{r,\mathbf{Q}}$, which can be neglected in leading order. Thus,   they can be approximated by $E_{\mathbf{k},+}-E_{\mathbf{p}',+}\simeq \hbar v_Dq_0$ provided that the linearized Dirac spectrum accurately describes the intermediate state at momentum $\mathbf{r}=\mathbf{p}'$.

Using Eq.\ (\ref{eq:matrixDirac0}), we find 
\begin{eqnarray}
    &&\langle \mathbf{p}',-;r,\mathbf{Q}| H_T |\mathbf{r}=\mathbf{p}',+;r,\mathbf{Q}\rangle
    \nonumber\\
    &&\qquad = \frac{w}{2} \left[1 +  e^{i\gamma_{\boldsymbol{p}'}}\right] \left[1 - e^{-i\gamma^\prime_{\boldsymbol{p}'}}\right] 
\end{eqnarray}
and 
\begin{eqnarray}
        &&\langle \mathbf{p}',-; \overline {\mathbf{r}=\mathbf{p}',-};\mathbf{k},+|
    H_T |\mathbf{k},+ \rangle
    \nonumber\\
    &&\qquad = \frac{w}{2} \left[1 +  e^{i\gamma_{\boldsymbol{p}'}}\right] \left[1 + e^{-i\gamma^\prime_{\boldsymbol{p}'}}\right] 
\end{eqnarray}
for the matrix elements of the tunneling Hamiltonian. Note that here, $\gamma_{\boldsymbol{p}'}$ and $\gamma^\prime_{\boldsymbol{p}'}$ are evaluated using the bond vectors of the respective layer. To compute the matrix elements of $H_\mathrm{intra}$, we use that it has only offdiagonal matrix elements in sublattice space. We can approximate the electron momenta as $\mathbf{k}\simeq \mathbf{K}$ and $\mathbf{p}' = \mathbf{r} \simeq \mathbf{K}'$. Using Eqs.\ (\ref{eq:bwfgra}) and (\ref{eq:helphintra}), 
we then have 
\begin{eqnarray}
    &&\langle \mathbf{r}=\mathbf{p}',+;r,\mathbf{Q}|
    H_\mathrm{intra} |\mathbf{k},+ \rangle \simeq - \delta_{\mathbf{Q},\mathbf{q}_0}
    \nonumber\\
    &&  \quad \times \frac{1}{2}\left\{ M_{\mathbf{K}',B;\mathbf{K},A}^r e^{i\gamma_{\mathbf{p}^\prime}}
    + [M_{\mathbf{K},B;\mathbf{K}',A}^r]^* e^{-i\gamma_\mathbf{k}} \right\} \quad
\end{eqnarray}
and 
\begin{eqnarray}
    &&\langle \mathbf{p}',-;r,\mathbf{Q}| H_\mathrm{intra} |\mathbf{p}',-; \overline {\mathbf{r}=\mathbf{p}',-};\mathbf{k},+\rangle \simeq  \delta_{\mathbf{Q},\mathbf{q}_0}\nonumber\\
    &&  \quad
    \times \frac{1}{2}\left\{ M_{\mathbf{K}',B;\mathbf{K},A}^r e^{i\gamma_{\mathbf{p}^\prime}} - [M_{\mathbf{K},B;\mathbf{K}',A}^r]^* e^{-i\gamma_\mathbf{k}} \right\}.\quad
\end{eqnarray}

Adding the amplitudes accounting for their relative minus sign, we find
\begin{widetext}
\begin{align}
    & \langle \mathbf{p}',-;r,\mathbf{Q}| H_\mathrm{intra} G_0 H_T - H_T G_0 H_\mathrm{intra} |\mathbf{k},+ \rangle =
    - \delta_{\mathbf{Q},\mathbf{q}_0} \frac{w}{4\hbar v_D q_0} \left[1 +  e^{i\gamma_{\boldsymbol{p}'}}\right]
    \nonumber\\
    & \quad \times
    \left\{
    \left[1 - e^{-i\gamma^\prime_{\boldsymbol{p}'}}\right]
 \left( M_{\mathbf{K}',B;\mathbf{K},A}^r e^{i\gamma_{\mathbf{p}^\prime}}
    + [M_{\mathbf{K},B;\mathbf{K}',A}^r]^* e^{-i\gamma_\mathbf{k}} \right) +  \left[1 + e^{-i\gamma^\prime_{\boldsymbol{p}'}}\right]
    \left( M_{\mathbf{K}',B;\mathbf{K},A}^r e^{i\gamma_{\mathbf{p}^\prime}} - [M_{\mathbf{K},B;\mathbf{K}',A}^r]^* e^{-i\gamma_\mathbf{k}} \right)\right\}
        \nonumber\\
    & \quad =
    - \delta_{\mathbf{Q},\mathbf{q}_0} \frac{w}{2\hbar v_D q_0} \left[1 +  e^{i\gamma_{\boldsymbol{p}'}}\right] \left\{
  M_{\mathbf{K}',B;\mathbf{K},A}^r e^{i\gamma_{\mathbf{p}^\prime}}
    - [M_{\mathbf{K},B;\mathbf{K}',A}^r]^* e^{-i\gamma_\mathbf{k}-i\gamma^\prime_{\boldsymbol{p}'}}  \right\}
    \nonumber\\
    & \quad = \delta_{\mathbf{Q},\mathbf{q}_0}
    \frac{w}{2\hbar v_D q_0}  \left[1 -  e^{-i(\theta_{\boldsymbol{\pi}'}-\theta/2)}\right] \left\{ M_{\mathbf{K}',B;\mathbf{K},A}^r e^{-i(\theta_{\boldsymbol{\pi}^\prime}-\theta/2)}
    + [M_{\mathbf{K},B;\mathbf{K}',A}^r]^* e^{i\theta_{\boldsymbol{\kappa}} + i\theta^\prime_{\boldsymbol{\pi}'}}  \right\}.
\end{align}
In the last step, we have approximated the phases in the Dirac approximation. Taking the absolute value, averaging over the Fermi circles, and noting that as a result of the average over $\theta_{\boldsymbol{\kappa}}$, the two terms contribute incoherently gives 
\begin{equation}
 \int \frac{d\theta_\mathbf{k}}{2\pi}\frac{d\theta_{\mathbf{p}^\prime}}{2\pi} |\langle \mathbf{p}',-;r,\mathbf{Q}|\mathcal{T}_\mathrm{inel}|\mathbf{k},+\rangle|^2  \to  \frac{w^2}{2(\hbar v_D q_0)^2} \left\{ |M_{\mathbf{K}',B;\mathbf{K},A}^r|^2 
    + |M_{\mathbf{K},B;\mathbf{K}',A}^r|^2   \right\}
\end{equation}
Inserting this into Eq.\ (\ref{eq:d2IdV2gen}) and accounting for phonon emission in the substrate, we find
\begin{equation}
    \frac{d^2\delta I_\mathrm{intra}}{dV_b^2}  = 
    G_\mathrm{incoh} \sum_r
    \delta(V_b-\hbar\omega_{r,\mathbf{q}_0}/e)
     \frac{\kappa_F^2\Omega}{(\hbar v_D q_0)^2}  \left\{ |M_{\mathbf{K}',B;\mathbf{K},A}^r|^2 
    + |M_{\mathbf{K},B;\mathbf{K}',A}^r|^2   \right\}
\end{equation}
Both matrix elements contribute equally, so that we find the result
\begin{equation}
    \frac{d^2\delta I_\mathrm{intra}}{dV_b^2}  = 
    G_\mathrm{incoh} \sum_r
    \delta(V_b-\hbar\omega_{r,\mathbf{q}_0}/e)(\kappa_F\ell_{r,\mathbf{q}_0})^2
     \frac{\beta^2\Omega_\mathrm{uc}}{(\hbar v_D q_0)^2}   \left|\sum_j\left[\boldsymbol{\epsilon}_{r,\mathbf{q}_0}^A
e^{-i\mathbf{q}_0\cdot \mathbf{e}_j} - \boldsymbol{\epsilon}_{r,\mathbf{q}_0}^B\right] \right|^2    
\label{eq:result_intra}
\end{equation}
\end{widetext}
The behavior in the limit of small twist angle, i.e., $\mathbf{q}_0\to 0$, depends on the type of phonon mode. For acoustic phonons, the dependences on $q_0$ of the last two factors on the right hand side compensate to yield a constant. At charge neutrality, 
$(\kappa_F\ell_r)^2\sim \omega_{r,\mathbf{q}_0}$ implies that the overall expression reduces linearly with decreasing $q_0\to 0$. Away from charge neutrality, the peak grows as $1/q_0$. For optical phonons, the last factor as well as $(\kappa_F\ell_r)^2$ remain constant for $q_0\to 0$. Thus, the peak grows as $1/q_0^2$ as the twist angle decreases both at and away from charge neutrality. 

We end this section by remarking that making the replacement derived in App.\ \ref{app:interlayer}, one can reproduce the contribution of acoustic phonons due to the interlayer electron-phonon interaction from Eq.\ (\ref{eq:result_intra}) for the intralayer coupling. 

\section{Conclusions}
\label{sec:conclusions}

While STM probes local tunneling in real space, the QTM exploits tunneling which is local in reciprocal space. This makes the QTM ideally suited to probing momentum-dependent dispersions. Elastic tunneling between twisted graphene layers requires voltages above a twist-angle dependent threshold. This opens a window at small voltages, in which the tunneling current in a QTM is purely inelastic and gives access to collective-mode dispersions. In this paper, we illustrated this modality by  developing a comprehensive and analytical theory for phonon spectroscopy in twisted graphene-graphene junctions. We showed that intra- and interlayer electron-phonon couplings give contributions to the inelastic tunneling current, and that the dominant contribution can come from various processes for a specific phonon mode, as summarized in Table \ref{tab:elph}. 

The intralayer electron-phonon coupling has two contributions at long wavelengths.  Phonon-induced modifications in the bond lengths can be described as a gauge field within the effective Dirac description of the graphene band structure. The conventional deformation potential associated with compressions and dilations of the lattice enters the Dirac equation as a scalar potential. It has been difficult to reliably extract corresponding coupling constants from experiment. Provided that the contribution of intralayer electron-phonon coupling contributes significantly in QTM measurements, we find that for small twist angles, the gauge coupling is much stronger for transverse acoustic modes, while the deformation potential couples only to longitudinal acoustic modes. Thus, QTM measurements may well resolve this issue in addition to providing direct access to coupling constants. 

Our considerations were limited to situations in which tunneling preserves the valley. In practice, rotating the tip by $\pi/3$ merely replaces the tip's $K$-valley by the $K'$-valley. This makes the twist-angle dependent tunneling invariant under $\pi/3$ rotations. Moreover, every phonon mode contributes twice at a particular twist angle, due to scattering between identical as well as opposite valleys. Both contributions become symmetric at a twist angle of $\pi/6$, making this a symmetry point of the measured spectra.   

Our analytical considerations focused on zero temperature, where only phonon emission contributes to the inelastic tunneling current. At finite temperatures, also phonon absorption contributes. We also did not consider higher-order umklapp processes. These do affect elastic scattering at specific angles, making their contribution to inelastic tunneling an interesting question. Both effects can be included by straightforward extension of the calculations presented here. 

Our considerations further neglected the finite size of the tunneling contact. A finite contact area limits the accuracy of momentum conservation in the tunneling process. This enables a temperature-dependent background current due to thermal phonons. The magnitude of this background depends sensitively on the contact shape. We can evaluate the tunneling amplitude for a finite contact area following the discussion of the scattering picture in Sec.\ \ref{sec:overview_scat}. While a rectangular contact would have independent Lorentzian-type broadenings of the $x$- and $y$-components of the momentum, these components are coupled in more general situations. For instance, one would expect just a single Lorentzian broadening for a weakly disordered contact. The background current is dominated by the tail of these broadening functions, which yields different behaviors in these two cases. 

In addition to phonon dispersions, QTM measurements also give access to momentum-resolved electron-phonon couplings. In view of the multiple inelastic tunneling processes, extracting this information from experiment must rely on theoretical results of the kind that we provide in this paper. Corresponding results promise to shed light on the nature of superconductivity and the linear temperature dependence of the resistivity in magic-angle twisted bilayer graphene. Phonon spectroscopy based on QTM measurements may finally bring us closer to understanding the origin of these phenomena, which may or may not originate from electron-phonon coupling.

\begin{acknowledgments}
Research at Weizmann, Yale, and Freie Universit\"at Berlin was supported by 
Deutsche Forschungsgemeinschaft through CRC 183 (project C02 and a Mercator Fellowship).  Research at Freie Universit\"{a}t Berlin was further supported by Deutsche Forschungsgemeinschaft through a joint ANR-DFG project (TWISTGRAPH). Research at Yale was supported by NSF Grant No.\ DMR-2410182 and by the Office of Naval Research (ONR) under Award No.\ N00014-22-1-2764.P.G.\ acknowledges support from the Severo Ochoa program for centers of excellence in R\&D (CEX2020-001039-S/AEI/10.13039/501100011033, Ministerio de Ciencia e Innovacion, Spain), from grant (MAD2D-CM)-MRR MATERIALES
AVANZADOS-IMDEA-NC, NOVMOMAT, as well as Grant PID2022-142162NB-I00 funded by MCIN/AEI/10.13039/501100011033. Work at Weizmann was  supported by the Leona M.\ and Harry B.\ Helmsley Charitable Trust grant, and the Rosa and Emilio Segre Research Award (S.I.) as well as an NSF-BSF award DMR-2000987, and the European Research Council (ERC) under grant HQMAT (Grant Agreement No. 817799) (E.B.).
\end{acknowledgments}

\appendix

\section{Interlayer electron-phonon coupling}
\label{app:interlayer}

It may seem accidental that for acoustic phonons, the contributions of first- and second-order perturbation theory are of the same order. Here, we rationalize this result by showing that the two kinds of electron-phonon couplings can be made to look similar by means of a gauge transformation. 

Focusing on the contribution of $T_0$, the interlayer tunneling can be written in the continuum limit as
\begin{align}
    \mathcal{H}_T &= \int d\mathbf{r} \psi_{t,\alpha}^\dagger(\mathbf{r}) T_0^{\alpha\beta}(\mathbf{r}) e^{i\mathbf{K}\cdot[\mathbf{u}(\mathbf{r}+\boldsymbol{\tau}_\alpha)-\mathbf{u}'(\mathbf{r}+\boldsymbol{\tau}_\beta)] } \psi_{b,\beta}(\mathbf{r})
    \nonumber\\
    & \qquad + \mathrm{h.c.}
\end{align}
The phonon displacements of the two layers contribute a phase to the interlayer tunneling, akin to a vector potential introducing a Peierls phase. This phase can be eliminated by a gauge transformation 
\begin{equation}
   \psi_{b,\beta}(\mathbf{r}) \to \psi_{b,\beta}(\mathbf{r}) e^{i\mathbf{K}\cdot \mathbf{u}'(\mathbf{r}+\boldsymbol{\tau}_\beta)} 
\end{equation}
and an analogous transformation for the upper layer. This is implemented by the unitary transformation
\begin{widetext}
\begin{equation}
    \mathcal{U} = \exp\{ - i\sum_\alpha \int d\mathbf{r} \mathbf{K}\cdot [\mathbf{u}(\mathbf{r}+\boldsymbol{\tau}_\alpha)
    \psi^\dagger_{t,\alpha}(\mathbf{r}) \psi_{t,\alpha}(\mathbf{r}) - \mathbf{u}'(\mathbf{r}+\boldsymbol{\tau}_\alpha)
    \psi^\dagger_{b,\alpha}(\mathbf{r}) \psi_{b,\alpha}(\mathbf{r})]  \}.
\end{equation}
Apart from eliminating the phase factor including the phonon displacements in $H_T$, this transforms the electronic Hamiltonian of the graphene layers. (An additional contibution due to the transformation of the phonon Hamiltonian is small in the parameter $c/v_D$.) We revert to a tight-binding description for convenience. Then, the unitary transformation becomes
\begin{equation}
    \mathcal{U} = \exp\left\{ - i\sum_\alpha \sum_\mathbf{R} \mathbf{K}\cdot \left[ \mathbf{u}(\mathbf{R}+\boldsymbol{\tau}_\alpha)
    c^\dagger_{t}(\mathbf{R}+\boldsymbol{\tau}_\alpha) c_{t}(\mathbf{R}+\boldsymbol{\tau}_\alpha) - \mathbf{u}'(\mathbf{R}'+\boldsymbol{\tau}_\alpha^\prime)
    c^\dagger_{b}(\mathbf{R}'+\boldsymbol{\tau}_\alpha^\prime) c_{b}(\mathbf{R}'+\boldsymbol{\tau}_\alpha^\prime) \right]   \right\}.
\end{equation}
The graphene Hamiltonian, say for the top layer, transforms into
\begin{equation}
    \mathcal{H} = - t^\parallel \sum_\mathbf{R} \sum_{\mathbf{e}_j}
    e^{i\mathbf{K}\cdot[\mathbf{u}(\mathbf{R})-\mathbf{u}(\mathbf{R}+\mathbf{e}_j)]}
    c^\dagger_{t}(\mathbf{R}+\mathbf{e}_j) c_{t}(\mathbf{R}) 
    + \mathrm{h.c.}
\end{equation}
\end{widetext}
Expanding the exponential to linear order in the atomic displacements, we find that the resulting electron-phonon coupling takes essentially the same form as the intralayer electron-phonon coupling of the top layer, except for the fact that it is phase shifted by $\pi/2$ due to the factor of $i$. The calculation for the bottom layer proceeds by complete analogy and generates a corresponding electron phonon coupling for the bottom layer. 

With this form of the interlayer electron-phonon coupling, we readily see that the results for intralayer and interlayer electron-phonon coupling are related by the replacement
\begin{equation}
    \frac{\partial t^\parallel}{\partial a} \leftrightarrow i t^\parallel \mathbf{K}.
\end{equation}
One can check that making this replacement in Eq.\ (\ref{eq:result_intra}) indeed reproduces the contribution of in-plane phonons to
Eq.\ (\ref{eq:ResultInter}) for the interlayer electron-phonon coupling. 

The vectorial nature of the interlayer coupling implies that it is strongly twist-angle dependent, with coupling to transverse phonons dominating over longitudinal phonons. This relation also makes it explicit that apart from this difference in their twist-angle dependence, the two types of electron-phonon couplings give contributions of the same order, when making the (approximate) replacement ${\partial t^\parallel}/{\partial a}\to { t^\parallel}/{ a}$. In particular, we expect that the parametric dependences on electron filling and phonon wave vector are identical in both cases. Moreover, the relative phase shift of $\pi/2$ implies that there are no interference terms between the two contributions. 

It is also interesting to note that the dependence of the electron-phonon coupling on the phonon momentum $\mathbf{Q}$ is seemingly different before and after the gauge transformation. Focusing on acoustic phonons, the coupling before the gauge transformation diverges as $1/\sqrt{Q}$ due to the linear dispersion. This dependence emerges from the zero-point amplitude of the phonon displacements. In contrast, after the transformation, there is an additional factor of $Q$ due to the difference in phonon displacements of neighboring graphene sites. Thus, the coupling behaves as $\sqrt{Q}$. This shows that the coupling is consistent with the expectation that it vanishes in the long-wavelength limit corresponding to near-uniform relative shifts of the top and bottom layers. Of course, in keeping with gauge invariance,
physical results are identical in both cases, with the singular $Q$-dependence emerging from the energy denominator after the gauge transformation. 


%

\end{document}